\documentclass[letterpaper,11pt]{article}
\pdfoutput=1

\usepackage{jheppub}
\usepackage{multirow}

\usepackage{subfig}
\usepackage{xspace}
\usepackage[countmax]{subfloat}

\usepackage{amssymb}
\usepackage{amsmath}
\usepackage{cancel}
\usepackage{color}
\usepackage{braket}
\usepackage{graphicx}
\usepackage{multirow}
\usepackage{verbatim}
\usepackage{amsthm}
\usepackage{slashed}
\usepackage{wasysym}
\usepackage{simplewick}
\usepackage{mathtools}
\usepackage{soul}
\usepackage{xspace}

\newcommand{\ecfop}[1]{\hat{\mathbf{E}}_{#1}}

\newcommand{\pythia}[1]{\textsc{Pythia\xspace #1}}

\def\nn{{\nonumber}}

\newcommand{\fastjet}[1]{\textsc{FastJet\xspace #1}}

\newcommand{\Eq}[1]{Equation~\eqref{#1}}

\def\zcut{z_{\text{cut}}}
\def\CSS{{\text{CSS}}}
\def\NP{{\text{NP}}}
\def\P{{\text{P}}}

\DeclareRobustCommand{\Sec}[1]{Sec.~\ref{#1}}
\DeclareRobustCommand{\Secs}[2]{Secs.~\ref{#1} and \ref{#2}}

\DeclareRobustCommand{\Fig}[1]{Fig.~\ref{#1}}

\DeclareRobustCommand{\Eq}[1]{Eq.~(\ref{#1})}

\def\be{\begin{equation}}
\def\ee{\end{equation}}

\newcommand{\ecf}[2]{e_{#1}^{(#2)}} 
\newcommand{\ecfnobeta}[1]{e_{#1}}

\allowdisplaybreaks[3]

\renewcommand{\P}{\mathrm{P}}       

  \newcount\hour \newcount\minute
  \hour=\time \divide \hour by 60 \minute=\time
  \newcommand{\todaytime}{\today \ -- \number\hour :\ifnum \minute<10 0\fi\number\minute}

\title{Convolved Substructure:\\Analytically Decorrelating Jet Substructure Observables}

\author[1,2]{Ian Moult,}
\author[3]{Benjamin Nachman}
\author[4]{and Duff Neill}
\affiliation[1]{Berkeley Center for Theoretical Physics, University of California, Berkeley, CA 94720, USA}
\affiliation[2]{Theoretical Physics Group, Lawrence Berkeley National Laboratory, Berkeley, CA 94720, USA}
\affiliation[3]{Physics Division, Lawrence Berkeley National Laboratory, Berkeley, CA 94720, USA}
\affiliation[4]{Theoretical Division, MS B283, Los Alamos National Laboratory, Los Alamos, NM 87545, USA}
\emailAdd{ianmoult@lbl.gov}
\emailAdd{bpnachman@lbl.gov}
\emailAdd{duff.neill@gmail.com}

\abstract{A number of recent applications of jet substructure, in particular searches for light new particles, require substructure observables that are decorrelated with the jet mass. In this paper we introduce the Convolved SubStructure (CSS) approach, which uses a theoretical understanding of the observable to decorrelate the complete shape of its distribution.  This decorrelation is performed by convolution with a shape function whose parameters and mass dependence are derived analytically. We consider in detail the case of the $D_2$ observable and perform an illustrative case study using a search for a light hadronically decaying $Z'$. We find that the CSS approach completely decorrelates the $D_2$ observable over a wide range of masses. Our approach highlights the importance of improving the theoretical understanding of jet substructure observables to exploit increasingly subtle features for performance.
}

\begin{document} 

\maketitle

\section{Introduction}\label{sec:intro}

Jet substructure is now playing a central role at the Large Hadron Collider (LHC), where it has provided a new set of powerful tools to search for physics beyond the Standard Model.  For example, jet substructure tools have been used to tag highly Lorentz boosted Standard Model bosons ($W/Z/H$), significantly improving searches for new high mass states (see e.g. \cite{ATLAS-CONF-2015-035,Aad:2015rpa,Aaboud:2016okv,Aaboud:2016trl,Aaboud:2016qgg,Aaboud:2017zfn,Aaboud:2017ahz,Aaboud:2017eta,Aaboud:2017itg,Aaboud:2017ecz,Khachatryan:2015bma,Khachatryan:2016cfa,Sirunyan:2016cao,Sirunyan:2017acf,Sirunyan:2017wto,Sirunyan:2017ukk,Sirunyan:2017bfa,Sirunyan:2017uhk}). With an ever improving understanding of jet substructure observables, these tools have now also been used to search for low mass resonances by directly studying the mass distribution of the tagged jets themselves. This has been applied both to the Standard Model search for $H\to b\bar b$ \cite{Sirunyan:2017dgc,CMS-PAS-HIG-17-010}, and to searches for new light $Z'$ bosons, deriving bounds in a previously unprobed region of parameter space \cite{CMS-PAS-EXO-17-001,Sirunyan:2017dnz,Sirunyan:2017nvi}.\footnote{For other recent bounds on this region see \cite{Mariotti:2017vtv}. } These searches represent an impressive advance in the sophistication of jet substructure techniques.

Unlike for high mass resonance searches, these low mass searches use the mass of the jet itself. This makes it important that the jet substructure observable used for tagging is independent of the mass of the jet. Otherwise, the cut on the tagging observable can significantly distort the jet mass spectrum, making it difficult to search for resonances. This was first highlighted in \cite{Dolen:2016kst}, where a procedure, termed DDT, was introduced to decorrelate the observable from the jet mass and $p_T$. More precisely, the DDT decorrelates the first moment of the observable. Due to the importance of this problem, several other groups have applied machine learning to develop tagging observables that are decorrelated with the jet mass and $p_T$ \cite{Shimmin:2017mfk,Aguilar-Saavedra:2017rzt}.

In parallel with experimental advances, there have been significant advances in the theoretical understanding of jet substructure observables,\footnote{For a review of recent advances in jet substructure, see \cite{Larkoski:2017jix}.} and a large number of calculations from first principles QCD \cite{Marzani:2017mva,Frye:2016aiz,Frye:2016okc,Banfi:2016zlc,Banfi:2015pju,Stewart:2013faa,Becher:2012qa,Feige:2012vc,Larkoski:2014wba,Dasgupta:2013ihk,Dasgupta:2013via,Dasgupta:2015lxh,Larkoski:2015kga,Frye:2017yrw,Jouttenus:2013hs,Dasgupta:2015yua,Hoang:2017kmk}. These calculations provide significant insight into the behavior of jet substructure observables, and have enabled advances in their sophistication, with many of the most important observables in current use arising out of analytic calculations. Recently an all orders factorization formula \cite{Larkoski:2017iuy,Larkoski:2017cqq} was derived for the groomed\footnote{By grooming we mean modified mass drop (MMDT) \cite{Dasgupta:2013ihk,Dasgupta:2013via} or soft drop \cite{Larkoski:2014wba} groomers, which for $\beta=0$ are equivalent.} $D_2$ observable \cite{Larkoski:2014gra,Larkoski:2015kga}, which is used extensively by ATLAS \cite{ATLAS-CONF-2015-035,Aad:2015rpa,Aaboud:2016okv,Aaboud:2016trl,Aaboud:2016qgg,Aaboud:2017zfn,Aaboud:2017ahz,Aaboud:2017eta,Aaboud:2017itg,Aaboud:2017ecz}. It was derived in soft-collinear effective theory (SCET) \cite{Bauer:2000yr,Bauer:2001ct,Bauer:2001yt,Bauer:2002nz,Rothstein:2016bsq} and its multi-scale extensions \cite{Bauer:2011uc,Larkoski:2014tva,Procura:2014cba,Larkoski:2015zka,Larkoski:2015kga,Pietrulewicz:2016nwo,Chien:2015cka}. This factorization allows for an understanding of the all orders perturbative and non-perturbative behavior of the observable. 

\begin{figure}
\begin{center}
\includegraphics[width=8.5cm]{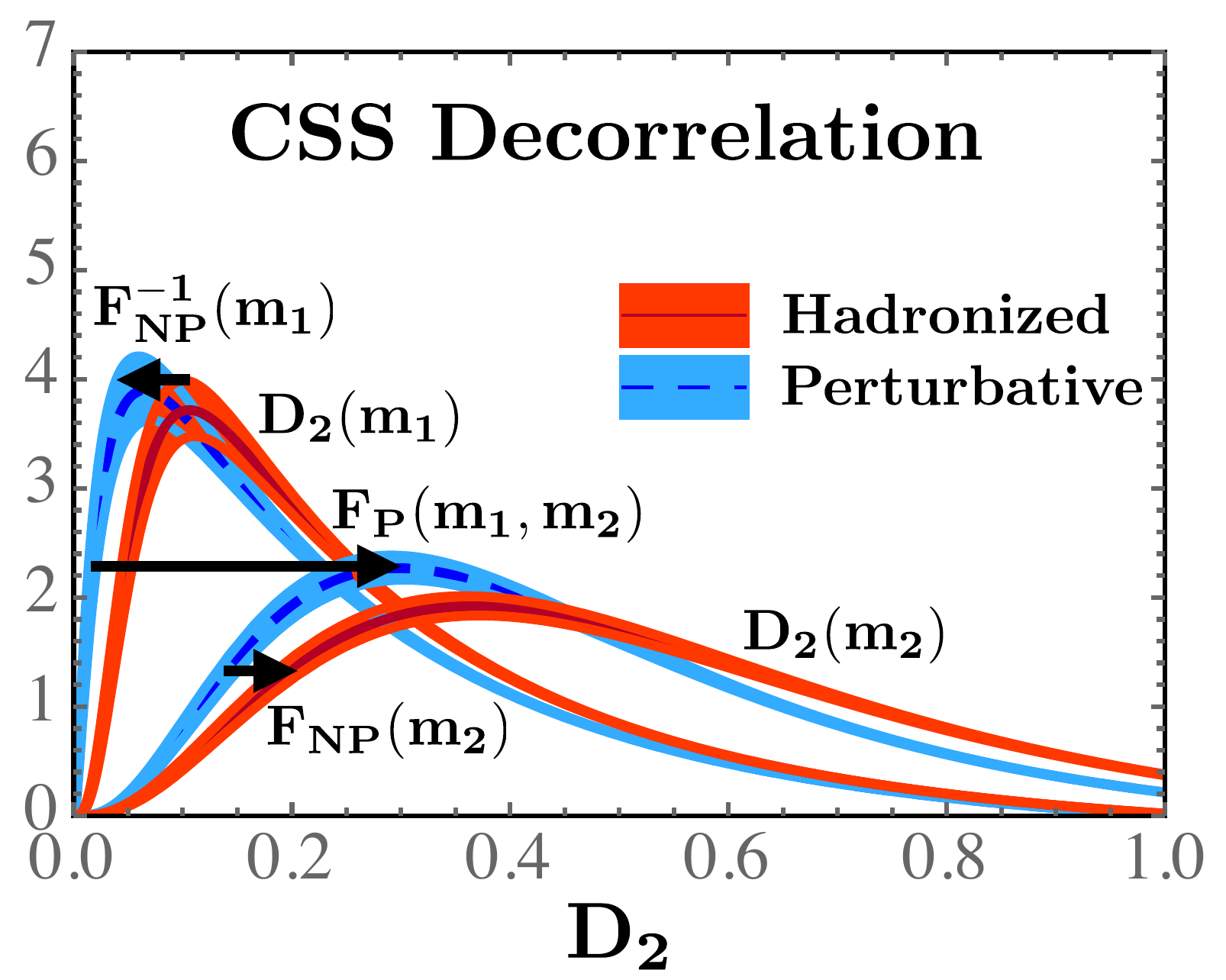}    
\end{center}
\caption{The evolution of a two-prong observable, taken here to be $D_2$, with the jet mass is governed by the corresponding evolution of its perturbative and non-perturbative components. Here $F_{\text{NP}}(\epsilon;m)$ encodes the effects of hadronization, while $F_{\text{P}}(\epsilon;m_1,m_2)$ is a perturbatively calculable function describing the mapping between the perturbative distributions at the masses $m_1$ and $m_2$ (They are technically defined as convolutions in $\epsilon$ as described in the text, which has been suppressed in the figure.). By combining these mappings we can completely decorrelate the observable by mapping it to a reference mass value.
}
\label{fig:intro}
\end{figure}

In this paper, we show how we can use an understanding of substructure observables to completely decorrelate them with the jet mass. In particular, we will show that the standard way of incorporating non-perturbative hadronization effects, namely convolution with a model shape function, motivates a simple way of performing the decorrelation: convolution with a function that maps the distribution at any mass to the distribution at a reference mass. We will call this approach to decorrelation Convolved SubStructure (CSS).\footnote{We note that CSS is also the common abbreviation for the pioneers of factorization, namely Collins, Soper and Sterman \cite{Collins:1981uk,Collins:1984kg,Collins:1985ue,Collins:1988ig,Collins:1989gx}. We find this fitting since our approach is based on a factorized understanding of the observable.}  The CSS approach naturally preserves the domain and normalization of the tagging observable, and allows a decorrelation of the complete shape of the observable, not just the first moment. The philosophy of our approach is slightly distinct from \cite{Aguilar-Saavedra:2017rzt,Shimmin:2017mfk}, namely it attempts to decorrelate a given standard observable, such as $D_2$ \cite{Larkoski:2014gra,Larkoski:2015kga}, or $N_2$ \cite{Moult:2016cvt}, using a theoretical understanding of that particular observable, and as such, is similar in spirit to the original DDT \cite{Dolen:2016kst}.  Indeed, we will show that the first moment of our approach reproduces the DDT, and therefore the CSS approach should be thought of as a systematic generalization of the DDT beyond the first moment.

A schematic depiction of our approach is shown in \Fig{fig:intro}. At a given mass, the distribution predicted by the factorization theorem for an observable such as $D_2$ is given as a convolution of a non-perturbative shape function \cite{Korchemsky:1999kt,Korchemsky:2000kp,Hoang:2007vb,Ligeti:2008ac} $F_{\text{NP}}(\epsilon;m)$ which encodes the effects of hadronization, with the perturbative distribution (here and throughout the text, $\epsilon$ will denote a dimensionless convolution variable, and $m$ denotes the mass). Both the perturbative distribution, as well as the non-perturbative shape function depend on the jet mass, and therefore both introduce correlations between the observable and the jet mass. However, with an understanding of these different functions, we can map the distribution at a given mass to a reference mass if we know both the non-perturbative shape function, $F_{\text{NP}}(\epsilon,m)$, as well as the mapping between the perturbative distributions, $F_{\text{P}}(\epsilon;m_1,m_2)$, which is a perturbatively calculable function. The end result is that we can derive a function, $F_\CSS(\epsilon;m_1, m_2)$, which completely decorrelates the observable by mapping it to a reference mass point.\footnote{Technically we map the \emph{graph} (the set points $\{(x,f(x)):x\in D\}$) of the observable to the \emph{graph} at a reference mass point.} This defines the CSS decorrelated $D_2$ observable:
\begin{align}
\frac{d\sigma^\CSS}{dD_2}=\int\limits_0^\infty d\epsilon\, F_\CSS(\epsilon;m_1, m_2)   \frac{d\sigma}{dD_2} \left(  D_2 - \epsilon \right)\,.
\end{align}
Here $\epsilon$ is a dimensionless convolution variable, and $m_1$ and $m_2$ denotes the masses that the function maps between.
The exact function can be determined through an understanding of both the perturbative and non-perturbative aspects of the distribution, namely,
\begin{align}\label{eq:conv_multiple_intro}
F_\CSS=F^{-1}_{\text{NP}} \otimes F_{\text{P}} \otimes F_{\text{NP}}\,,
\end{align}
where $\otimes$ denotes convolution. This combination of mappings is shown schematically in \Fig{fig:intro}.
While it is of course trivial that such a function exists, the simple structure of the observable enables us to provide a simple analytic form for the function $F_\CSS$, allowing for a fast numerical implementation, as well as an understanding of how it scales with $m_1$ and $m_2$. Furthermore, the function $F_\CSS$ can be systematically improved starting from this initial function, using an expansion in orthogonal polynomials, as developed in \cite{Ligeti:2008ac}\footnote{The perturbative distribution can of course be calculated, while the non-perturbative contribution must currently be modeled. However, due to the structure of the factorization theorem for the tagging observable, one can confidently predict the scaling of the non-perturbative corrections with the jet mass (that is, their contributions to the moments of the distribution) in a systematically improvable manner, thus fixing the functional form of the jet mass dependence in the expansion of the shape function with respect to the orthogonal polynomials (specifically, generalized Laguerre polynomials). }.

An outline of this paper is as follows. In \Sec{sec:correl} we discuss the sources of correlation between a two-prong substructure observable such as $D_2$, and the jet mass, treating both the perturbative and non-perturbative aspects of this correlation, and we show that in both cases they can be modeled using shape functions. Furthermore, we analytically derive the mass scaling of the shape function parameters. In \Sec{sec:hm} we discuss how we can use this understanding to decorrelate jet substructure observables using shape functions, and introduce the CSS approach. We then illustrate concretely how the decorrelation can be done in practice. In \Sec{sec:ph} we perform a brief study, illustrating the effectiveness of the decorrelation procedure for $Z'\to q\bar q$. We conclude in \Sec{sec:conc}.

\section{Correlation with Mass for Jet Substructure Observables}\label{sec:correl}

In this section we discuss the sources of correlation between a two-prong observable, such as $D_2$, and the jet mass (for brevity, we will not always explicitly say groomed jet mass, although we always work with groomed observables), to illustrate how these correlations arise. In \Sec{sec:NP}, we discuss the dependence of non-perturbative physics on jet mass, and introduce the modeling of hadronization effects using shape functions. In \Sec{sec:P} we discuss perturbative sources of correlation, and show that they can also be well captured by a simple shape function.

In this paper we will consider the concrete example of the $D_2$ observable, for which a factorization formula is known \cite{Larkoski:2017iuy,Larkoski:2017cqq}. This allows us to make precise statements about the perturbative and non-perturbative behavior of the observable. The $D_2$ observable is defined in terms of the energy correlation functions \cite{Larkoski:2013eya}
\begin{align}\label{eq:ppe2}
\ecf{2}{\beta}&=\frac{1}{p_{TJ}^2}\sum_{i<j\in J} p_{Ti} p_{Tj}R_{ij}^\beta\,, \\
\ecf{3}{\beta}&=\frac{1}{p_{TJ}^3}\sum_{i<j<k\in J} p_{Ti} p_{Tj}p_{Tk}R_{ij}^\beta R_{ik}^\beta R_{jk}^\beta\,,
\end{align}
as \cite{Larkoski:2014gra,Larkoski:2015kga}
\begin{equation}
D_2^{(\beta)} = \frac{\ecf{3}{\beta}}{(\ecf{2}{\beta})^3}\,.
\end{equation}
Here $R_{ij}$ is the distance between particles $i$ and $j$ in the pseudorapidity-azimuth plane, and $\beta>0$ is an angular weighting parameter whose typical value is $\beta=1$ or $\beta=2$. For notational simplicity we will often drop the angular exponent, writing the observable simply as $D_2$. 
For a jet with two prong substructure we have $D_2 \ll 1$, while for a more standard QCD jet without a resolved substructure $D_2 \sim 1$.  

\subsection{Non-Perturbative Effects}\label{sec:NP}

Jet substructure observables are sensitive to low scales within a jet, and are therefore naturally susceptible to non-perturbative effects. Non-perturbative contributions can arise both from the underlying event (UE), as well as from the standard hadronization process within the jet. In \cite{Larkoski:2017iuy}, it was shown that due to the grooming procedure, non-perturbative effects from the underlying event are negligible. We will therefore neglect them in what follows.

Using the factorization formula for the $D_2$ observable derived in \cite{Larkoski:2017iuy,Larkoski:2017cqq}, it can be shown that the dominant non-perturbative effects from hadronization are captured by a collinear-soft function
\begin{align}\label{eq:cs_func}
\hspace{-1cm}C_{si}(\ecfnobeta{3})&=\text{tr}\langle 0|T\{Y_i\}\delta (\ecfnobeta{3}-\ecfop{3}) \Theta_{\text{SD}}\bar{T}\{Y_i\} |0\rangle\,.
\end{align}
Here the $Y_i$ are products of Wilson lines along the subjet directions, $T$ and $\bar T$ denote time and anti-time ordering respectively. The measurement function and soft drop constraints are implemented by the energy flow operators $\ecfop{3}$ and $\Theta_{\text{SD}}$, whose exact form is not relevant for the current discussion. These operators can be written in terms of the energy-momentum tensor \cite{Sveshnikov:1995vi,Korchemsky:1997sy,Lee:2006nr,Bauer:2008dt}. Importantly, due to the application of the grooming algorithm, the collinear-soft function, and hence the non-perturbative hadronization corrections, depend only on the color structure of the jet itself, and not on the color structure of the global event, making them a property of the observable.

While the collinear-soft function in \Eq{eq:cs_func} can be calculated perturbatively, it is currently not possible to calculate it non-perturbatively. Instead, a functional parametrization of the non-perturbative matrix element, which is referred to as a shape function, $F_\NP$, is used \cite{Korchemsky:1999kt,Korchemsky:2000kp,Hoang:2007vb,Ligeti:2008ac}. Shape functions  have been used in a variety of contexts in jet physics \cite{Abbate:2010xh,Stewart:2014nna,Hoang:2014wka,Larkoski:2015kga,Hoang:2017kmk,Larkoski:2017iuy}. For the particular case of $D_2$, this allows the non-perturbative $D_2$ distribution to be written as a convolution of the perturbative distribution and the shape function
\begin{align}
\frac{d\sigma_{\text{NP}}}{dD_2}=\int\limits_0^\infty dx\, \tilde F_\NP(x)   \frac{d\sigma}{dD_2} \left(  D_2 -\frac{x}{m_J \zcut^{3/2}} \right)\,.
\end{align}
The scalings entering this expression are determined by the scalings of the collinear-soft function in \Eq{eq:cs_func}, and were derived in \cite{Larkoski:2017iuy,Larkoski:2017cqq}. We will take our model shape function to have the simple functional form\footnote{This functional form is that of a Gamma distribution. Amusingly, we note that these are the maximally entropic distributions with a fixed first moment and first logarithmic moment.}
\begin{align}
\label{eq:shape_func_simple}
\tilde F_\NP(x;\alpha, \Omega_D)= \left(   \frac{\alpha}{\Omega_D}   \right)^\alpha   \frac{1}{\Gamma(\alpha)}  x^{\alpha-1}  e^{-\frac{\alpha x}{\Omega_D}}\,.
\end{align}
This function has a first moment $\Omega_D\sim \Lambda_{\text{QCD}}$, is normalized to unity, and we may think of this specific shape function as but the first term in an orthogonal expansion which specifies the non-perturbtive corrections to all moments of the distribution, where we have truncated to specifically fix only the first moment. Here $\alpha$ is a parameter, which specifies the functional form. We will choose $\alpha$ such that the function vanishes as $x\to 0$. We find that $\alpha=2$-$3$ provides a good description of the non-perturbative correction. Since the dominant effect is a shift of the first moment, which is fixed, it is only at small value of $D_2$ that there is dependence on $\alpha$.  The physical interpretation of this function is that it smears the energies within the jet at the scale $ \Lambda_{\text{QCD}}$. In certain cases universal properties of the first moment of shape functions can be proven \cite{Lee:2007jr,Lee:2006fn}. These moments, as well as higher moments have been extracted from event shape data, for example from the thrust event shape \cite{Abbate:2012jh}.

\begin{figure}
\begin{center}
\subfloat[]{\label{fig:nonpert_before}
\includegraphics[width=7cm]{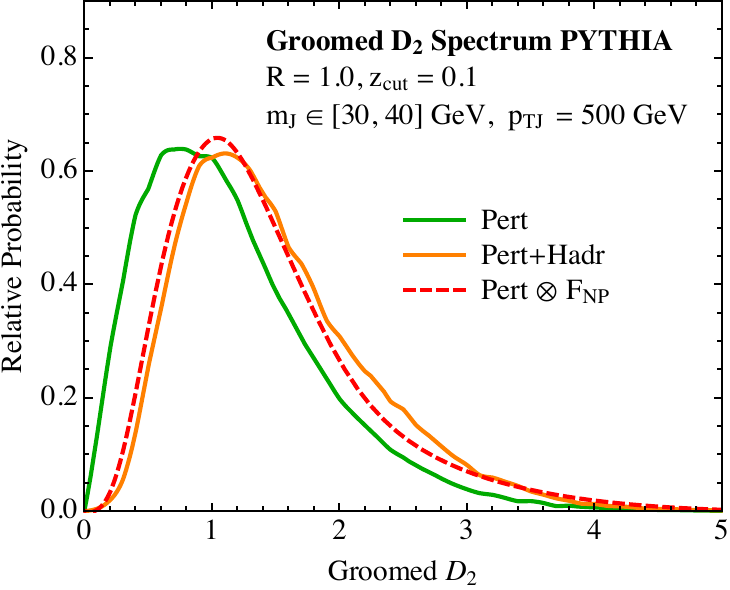}    
}\qquad
\subfloat[]{\label{fig:nonpert_scaling}
\includegraphics[width=7cm]{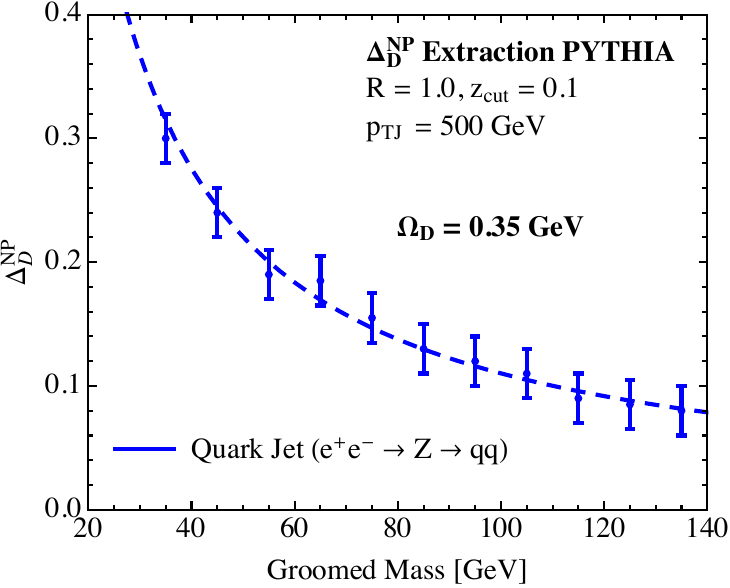}
}
\end{center}
\caption{The shift of the $D_2$ distribution due to hadronization. (a) The perturbative and hadronized distributions as found in \pythia{}, and as modeled using the non-perturbative shape function described in the text.  (b) The dependence of the non-perturbative shift, $\Delta_D^\NP$ as a function of the groomed mass, which introduces a source of correlation of the $D_2$ distribution with the groomed jet mass.
}
\label{fig:nonpert}
\end{figure}

Ref.~\cite{Larkoski:2017iuy} studied the non-perturbative shape parameter $\Omega_D$, and found
\begin{itemize}
\item $\Omega_D$ is independent of the quark or gluon nature of the jet.
\item The scaling predicted by \Eq{eq:cs_func}, namely that the non-perturbative shift in the distribution is inversely proportional to the mass, is well respected in parton shower Monte Carlo simulations.
\end{itemize}
In \Fig{fig:nonpert}, we show the effects of hadronization on the $D_2$ observable found in \pythia{}, and as modeled using the shape function of \Eq{eq:shape_func_simple}. We see that the simple shape function reproduces quite well the effects of the hadronization.

Although it is conventional to work with a shape function parameter that has mass dimension $1$, such as $\Omega_D$, for our purposes it will be convenient to introduce the dimensionless shift in the first moment of the $D_2$ distribution, which we denote $\Delta_D^\NP$. For the case of the non-perturbative hadronization corrections, we have the relation
\begin{align}
\Delta_D^\NP =\frac{\Omega_D}{m_J \zcut^{3/2}}.
\end{align}
When using the dimensionless variable, we use the shape function
\begin{align}
F_\NP(\epsilon;\alpha, \Delta_D)= \left(   \frac{\alpha}{\Delta_D}   \right)^\alpha   \frac{1}{\Gamma(\alpha)}  \epsilon^{\alpha-1}  e^{-\frac{\alpha \epsilon}{\Delta_D}}\,.
\end{align}
which is the same functional form as in \Eq{eq:shape_func_simple}, but we have dropped the tilde to emphasize that the dimension of the argument has changed.
The dependence of $\Delta_D^\NP$ as extracted from \pythia{} is shown in \Fig{fig:nonpert_scaling}, as well as a fit for the non-perturbative parameter $\Omega_D$. To extract this scaling, we have fit the shift parameter in the tail region of the distribution, where we expect that a shift of the distribution is valid. The uncertainties represent a conservative estimate due to the fact that the precise region in which one should be performing the fit is not always clear. The strong dependence on the mass of the jet is clearly visible, which introduces a non-perturbative correlation between the $D_2$ distribution and the jet mass. It is also important to note that the shift $\Delta_D^\NP$ is dependent only on $m_J$, and not on $p_T$, as can be derived from the factorization formula \cite{Larkoski:2017iuy,Larkoski:2017cqq}. This simplification is only true for groomed distributions.

Inverting the logic of this section, if we are able to transform between the perturbative and non-perturbative distributions using a convolution with a simple function, this also implies that we can perform the deconvolution to obtain the perturbative distribution. Doing this would remove the correlation of the $D_2$ distribution with the jet mass arising from hadronization corrections. However, to completely decorrelate the distribution, we also need to understand how to decorrelate the perturbative distributions, which can also depend on the jet mass. This will be addressed in \Sec{sec:P}. In \Sec{sec:hm} we will then give a numerically simple way of performing the decorrelation via convolution. 

\subsection{Perturbative Effects}\label{sec:P}

\begin{figure}
\begin{center}
\subfloat[]{\label{fig:pert_before}
\includegraphics[width=7cm]{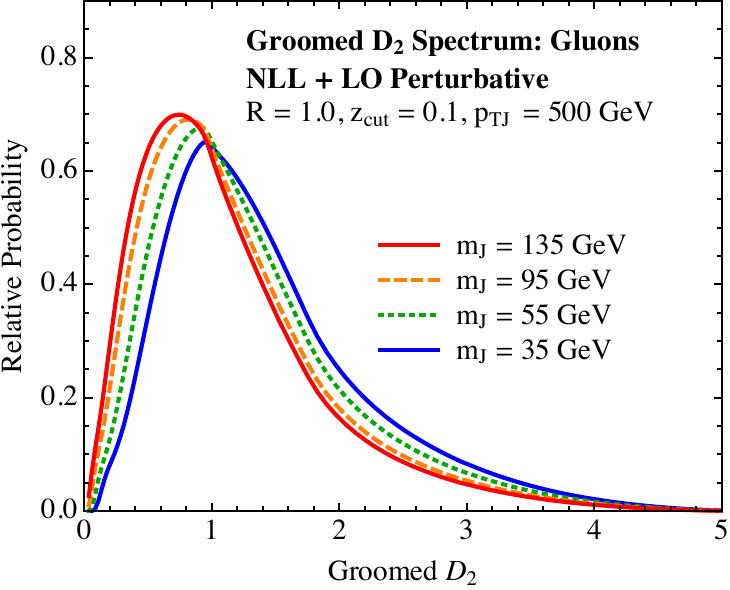}    
}\qquad
\subfloat[]{\label{fig:pert_after}
\includegraphics[width=7cm]{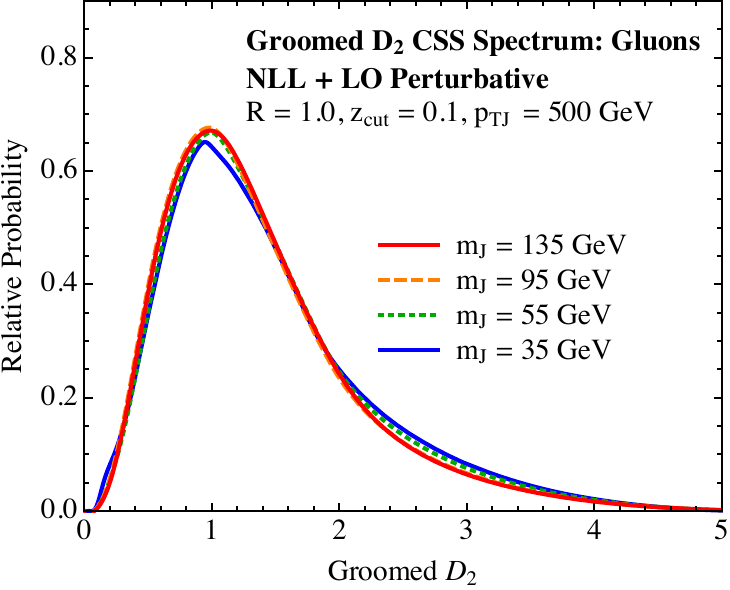}
}
\end{center}
\caption{The decorrelation of the perturbative $D_2$ spectrum. In (a) we show the perturbative groomed $D_2$ spectrum as a function of the jet mass, and in (b) we show the decorrelated $D_2$ spectrum. The movement of the distribution in (a) as the mass is varied is largely eliminated by the decorrelation procedure in (b).
}
\label{fig:pert}
\end{figure}

\begin{figure}
\begin{center}
\subfloat[]{\label{fig:pert_extracted_mean}
\includegraphics[width=6.9cm]{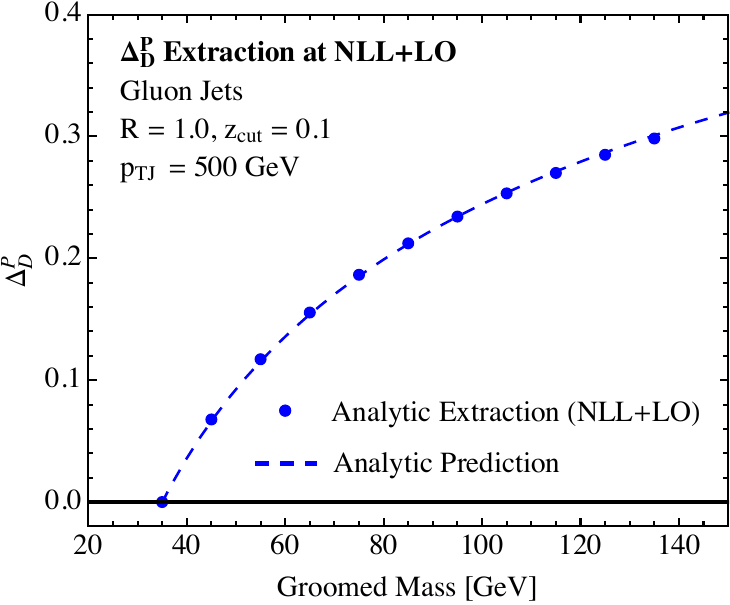}    
}\qquad
\subfloat[]{\label{fig:pert_extracted_logmean}
\includegraphics[width=7.2cm]{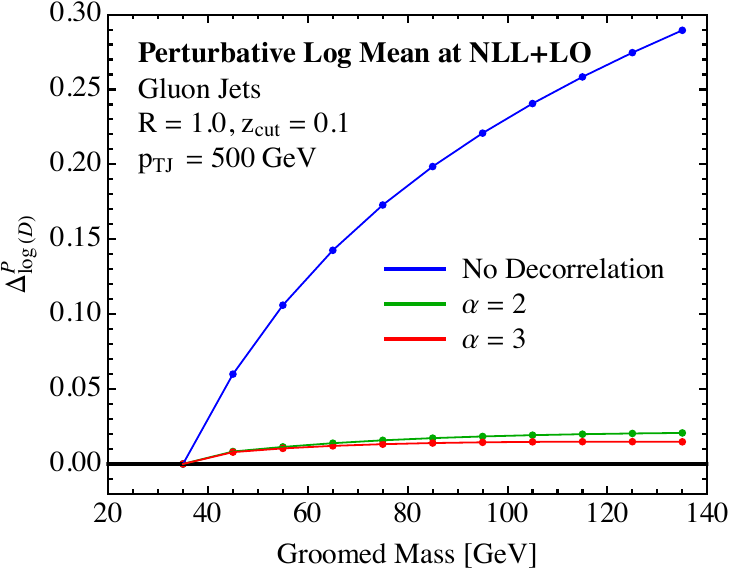}
}
\end{center}
\caption{(a) The shift in the mean of the perturbative distribution computed using the NLL$+$LO result, along with the analytic prediction described in the text. (b) The shift in the log mean for both the standard and CSS decorrelated distributions for different values of the $\alpha$ parameter for the shape function.
}
\label{fig:pert_extracted}
\end{figure}

In addition to a dependence of the hadronization corrections on the jet mass, there is also a dependence of the perturbative $D_2$ distribution on the jet mass that introduces a further correlation between the $D_2$ distribution and the jet mass. Unlike the hadronization correction, where only the scaling of the hadronization corrections as a function of the jet mass is calculable, the perturbative distribution can be calculated to a given accuracy, and hence the complete dependence of the distribution on the jet mass can be understood. In \Fig{fig:pert_before} we show a plot of the perturbative $D_2$ distribution at next-to-leading logarithm matched to leading order $1\to 3$ splitting functions in the large $D_2$ region in order to reproduce the correct endpoint behavior. In the figures this accuracy is referred to as NLL+LO. See \cite{Larkoski:2017cqq} for a more detailed discussion of the order counting. Here the $H\to gg$ process was used to produce gluon jets. We can see that there is a mild, but non-negligible dependence on the jet mass within the peak region. A more quantitative measure, $\Delta_D^P$, the shift in the mean relative to the distribution at $m=35$ GeV is shown in \Fig{fig:pert_extracted_mean}. This is only a small effect for the groomed $D_2$, which has a fixed endpoint at $1/(2\zcut)$, independent of the jet mass. It is ultimately this fact that leads to a large degree of stability of the distribution. For the ungroomed $D_2$ distribution the endpoint depends strongly on the jet mass so that the distribution displays a much more complicated dependence on the jet mass.

Following the logic of the previous section, if we understand the form of the correlation between the $D_2$ distribution and the jet mass, we can also remove this correlation. Motivated by the implementation of the shape function for the non-perturbative contribution, we can also attempt to decorrelate the perturbative component of the distribution by convolving with a function which takes the perturbative distributions to some reference value. Since the mean of the $D_2$ distribution increases with decreasing mass, to decorrelate by convolution with a simple shape function, we will always use as a reference mass value the lowest mass value of interest.  Namely, we write
\begin{align}
\frac{d\sigma}{dD_2}(\epsilon_1;m_2)=\int\limits_0^\infty d\epsilon\, F_P(\epsilon;m_1,m_2)   \frac{d\sigma}{dD_2} \left(  \epsilon_1 - \epsilon;m_1 \right)\,,\qquad m_2 < m_1.
\end{align}
Here we have made explicit the mass dependence of the functions, which is separated from the argument of the function by a semi-colon.
The fact that such a (possibly singular) function exists is trivial, and it can be determined by division in Laplace or Fourier space (i.e. by deconvolution). Furthermore, this function is (in principle) exactly calculable from the factorization theorem, given predictions for the perturbative $D_2$ distribution at any given accuracy at any jet mass. However, to have a reasonable prediction for the $D_2$ distribution requires a matched calculation. This implies that results for the distribution are necessarily numerical instead of analytic, making it difficult to understand the deconvolution analytically. We would therefore like to find a simple function that provides a good approximation to the exact result. 

Although we cannot analytically predict the exact shape function (in a practical way), we can use our analytic NLL$+$LO result to compute moments of the perturbative distribution. We expect that the dominant effect of the correlation between the $D_2$ observable and the jet mass will be a shift of the first moment, as can be seen from \Fig{fig:pert_before}. The shift in the mean relative to the distribution at $m=35$ GeV, $\Delta_D^P$, is shown in \Fig{fig:pert_extracted_mean}. The shift in the first moment of the distribution arises due to the renormalization group evolution of the functions appearing in the factorization theorem of Refs.~\cite{Larkoski:2017iuy,Larkoski:2017cqq}. We can therefore write the shift in the first perturbative order as
\begin{align}\label{eq:pert_shift_analytic}
\Delta_D^P= \gamma_D \int\limits_{m_2}^{m_1}  d\mu \frac{\alpha_s(\mu)}{\mu}+...\,,
\end{align}
where $\gamma_D$ is a constant, which we extract from our calculation of the distribution at two mass points. The prediction from this functional form is shown in the dashed line in \Fig{fig:pert_extracted_mean}, which provides an excellent description of the numerical results at many other values of the jet mass, confirming the perturbative evolution of the first moment. 

To perform the perturbative decorrelation, we will use as the base decorrelation function the functional form of \Eq{eq:shape_func_simple}. Since we can analytically predict the shift $\Delta_D^P$, we can use this function to exactly decorrelate the mean. However, by tuning the angular exponent, with the mean fixed, we can further attempt to decorrelate the complete shape of the distribution. The value of $\alpha$ can be extracted by decorrelating the log-mean of the distribution, which can be computed analytically from our NLL$+$LO calculation. The evolution of the log mean with mass is shown in \Fig{fig:pert_extracted}, both without decorrelation, and after decorrelation using the function of \Eq{eq:shape_func_simple} for several values of $\alpha$. We find that for $\alpha$ in the range of $\alpha=2$-$3$, we have good decorrelation of the log mean. Furthermore, it is quite insensitive to the exact value of $\alpha$ used, which shows that the correlation is dominated by a shift in the mean. The decorrelation of the full distribution for $\alpha=2.4$ is shown in \Fig{fig:pert_after}. As compared with \Fig{fig:pert_before}, we see a good decorrelation of the full shape of the distribution. This shows that the dependence of the $D_2$ observable on the mass is in fact remarkably simple, being driven by a shift in the first moment captured by \Eq{eq:pert_shift_analytic}, with deviations from this to account for the behavior at the endpoints being captured by the simple class of functions in \Eq{eq:shape_func_simple}.

We conclude this section by emphasizing that this analysis could be improved by iteratively building up a shape function starting from the base function of \Eq{eq:shape_func_simple} using an expansion in orthogonal functions, as has been done in \cite{Ligeti:2008ac}, requiring all moments to be decorrelated exactly. However, for our purposes we will find that the simple function of \Eq{eq:shape_func_simple} works extremely well, as will be illustrated in our case study in \Sec{sec:ph}.

\section{Convolved Substructure}\label{sec:hm}

Motivated by the above observation that both the perturbative and non-perturbative components of the distribution can be decorrelated using simple shape functions, we propose that we can use shape functions as an efficient way to completely decorrelate two-prong substructure observables by mapping them to reference mass. This is what we will call the Convolved Substructure, or CSS procedure. Since the shape functions used in hadronization are typically used to shift the distribution to a larger value, for the $D_2$ observable, we will also choose the reference mass to be the lowest mass of interest, ensuring that the shift in the mean required for decorrelation is positive. 

We define the CSS decorrelated $D_2$ observable by
\begin{align}\label{eq:D2_CSS_def}
\frac{d\sigma^\CSS}{dD_2}=\int\limits_0^\infty d\epsilon\, F_\CSS(\epsilon)   \frac{d\sigma}{dD_2} \left(  D_2 - \epsilon \right)\,.
\end{align}
Here $F_\CSS$ is an as of yet unspecified function with unit norm. While we have used the specific example of $D_2$, this approach should apply much more generally, however, we expect that it will only be for IRC safe observables with sufficiently favorable factorization properties that analytic scalings for the $F_\CSS$ function can be derived. Within this subset of observables, we believe that this represents a completely general and efficient way of performing the decorrelation. Unlike previously proposed analytic approaches, it aims to decorrelate all moments of the distribution, and naturally preserves the domain and norm of the distribution. Furthermore, motivated by the success in describing non-perturbative corrections using a simple basis of functions \cite{Ligeti:2008ac}, we will show that we can choose a simple analytic form of the function $F_\CSS$ as the initial approximation. Further improvements can be systematically added, if needed.

It is also interesting to see that this approach includes as a special case the standard DDT, which is a shift of the first moment.
Performing a Taylor expansion for a small shift, we have
\begin{align}\label{eq:taylor_CSS}
\frac{d\sigma^\CSS}{dD_2}&\simeq\int\limits_0^\infty d\epsilon\, F_\CSS(\epsilon)   \frac{d\sigma}{dD_2} \left(  D_2 \right)   - \int\limits_0^\infty d\epsilon\, F_\CSS(\epsilon) \epsilon \frac{d}{dD_2}  \frac{d\sigma}{dD_2} \left(  D_2 \right)  \nn \\
&\simeq \frac{d\sigma}{dD_2} \left(  D_2-\Delta_D \right)\,,
\end{align}
where $\Delta_D$ is the first moment of the function $F_\CSS$,
\begin{align}
\Delta_D=\int\limits_0^\infty d\epsilon\, F_\CSS(\epsilon) \epsilon\,.
\end{align}
This reproduces (a constrained form of) the DDT, which decorrelates the first moment. We note that while the DDT procedure was originally introduced as a shift which decorrelates the first moment of the distribution, it has since been generalized to decorrelate, for example, the background efficiency at a given cut. Nevertheless, it can still only decorrelate a single chosen moment of the distribution. We will re-emphasize this point in our numerical comparisons in \Sec{sec:ph}. Note that when used for incorporating non-perturbative effects, the linear shift applies in a particular region of the distribution, but the full shape function is needed at small values. We will see in \Sec{sec:practical} that this is also true when used for decorrelation, with the full convolution reducing to a linear shift throughout most of the distribution, and the full non-linear nature of the function only becoming relevant near the endpoints of the distribution.

\begin{figure}
\begin{center}
\includegraphics[width=8.5cm]{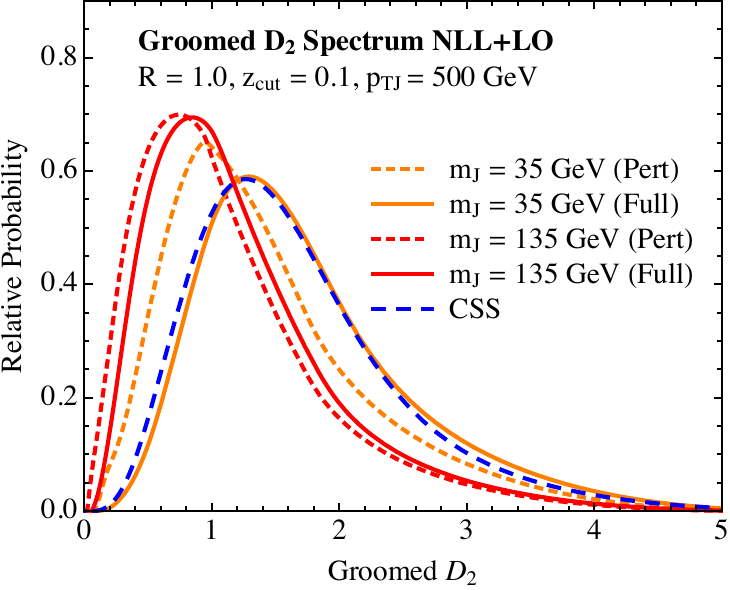}    
\end{center}
\caption{The implementation of the CSS decorrelation on our analytic NLL$+$LO calculation, using the first moment shift determined analytically from \Eq{eq:total_moment}. Perturbative distributions at the different mass values are shown in small-dashed, while the full distributions are shown in solid. The CSS result decorrelated to $m_J=35$ GeV is shown in dashed blue. It involves decorrelating both the perturbative and non-perturbative evolution, as can be seen from the different curves.
}
\label{fig:CSS_full}
\end{figure}

The exact function $F_\CSS$ to shift from the mass $m_1$ to a reference mass $m_2$, with $m_2<m_1$, can be written as
\begin{align}\label{eq:conv_multiple}
F_\CSS(\epsilon;m_1,m_2)=F^{-1}_{\text{NP}}(\epsilon;m_1) \otimes F_{\text{P}}(\epsilon;m_1,m_2) \otimes F_{\text{NP}}(\epsilon;m_2)\,,
\end{align}
as was illustrated in \Fig{fig:intro}. Here the $\otimes$ denotes convolution in the variable $\epsilon$, and the inverse denotes an inverse in the convolutional sense (i.e. a deconvolution). Instead of performing the decorrelation in this form, we will simplify our discussion and use a single effective function. This can certainly be improved, however, we will already find that with a single function we will find an excellent decorrelation. We will use the decorrelation function of the previous section, namely\footnote{That the final convolution in \Eq{eq:conv_multiple} can be approximated by a single function of the same form can be understood by looking at the functional form in Laplace space, where these functions take the form of rational functions to the power $\alpha$ using the first term in the expansion for $F_{\CSS}$. Due to the inverse convolution appearing in \Eq{eq:conv_multiple}, the Laplace transform of the convolution of the three functions has the same polynomial degree as the Laplace transform of a single such function.} 
\begin{align}
\label{eq:shapefunc}
F_\CSS(\epsilon;\alpha, \Delta_D)= \left(   \frac{\alpha}{\Delta_D}   \right)^\alpha   \frac{1}{\Gamma(\alpha)}  \epsilon^{\alpha-1}  e^{-\frac{\alpha \epsilon}{\Delta_D}}\,.
\end{align}
With this parametrization, we have that the first moment is $\Delta_D$ for all values of $\alpha$, but we allow for a general power law behavior as $x\to 0$, specified by $\alpha$. When considering a full example at the LHC, we will find that a value of $\alpha$ slightly larger than two will give an excellent fit. Taking the first moment of \Eq{eq:conv_multiple}, we find that
\begin{align}\label{eq:total_moment}
\Delta_D(m_1, m_2)=\Delta_D^\NP(m_2)-  \Delta_D^\NP(m_1) + \Delta_D^P(m_1,m_2)\,.
\end{align}
Again, we assume that the reference mass that we are shifting the distributions to, namely $m_2$, satisfies $m_2<m_1$.
In \Secs{sec:NP}{sec:P} we have used the factorization formula for the $D_2$ observable derived in \cite{Larkoski:2017iuy,Larkoski:2017cqq} to predict the mass dependence of both the perturbative, $\Delta_D^P$, and non-perturbative, $\Delta_D^\NP$, moments appearing in \Eq{eq:total_moment}. In principle, the exact values of the moments can be extracted for given processes and observables, by studying the distributions with and without hadronization, as was done above. 

The decorrelation using this procedure on our NLL$+$LO calculation is shown in \Fig{fig:CSS_full}, which shows both the perturbative and non-perturbative distributions, as well as the final CSS curve, and can be viewed as an analytic realization of the strategy outlined in \Fig{fig:intro}. Good, but not perfect decorrelation is observed, and we will see in \Sec{sec:ph} that the decorrelation procedure seems to work even better in \pythia{} than for the analytic example shown here.\footnote{There is also a tradeoff between exactly reproducing the mean and accurately capturing other aspects of the shape.} For ease of applicability, we find it more convenient to give a formula for $\Delta_D(m_1, m_2)$, with two constants that can be directly extracted by fitting the decorrelation at several points, as will be demonstrated in a practical example in \Sec{sec:ph}. Using our understanding of the functional dependence on the jet mass for both the perturbative and non-perturbative contributions to the moment discussed in \Secs{sec:NP}{sec:P}, we have the general form of the moment for the CSS approach as 
\begin{align}\label{eq:total_moment_formula}
\Delta_D(m_1, m_2)&=c_{\NP}\left(\frac{1}{m_2}-\frac{1}{m_1}   \right) +c_P  \int\limits_{m_2}^{m_1}  d\mu \frac{\alpha_s(\mu)}{\mu}\,, \\
&\simeq c_{\NP}\left(\frac{1}{m_2}-\frac{1}{m_1}   \right) +\tilde c_P \log\left(\frac{m_1}{m_2}  \right)\,,
\end{align}
where the second line is an approximation that is good for most numerical purposes. Again, we emphasize that the reference mass, $m_2$ is taken to satisfy $m_2<m_1$, so that this shift is positive. Here the $c_\NP$, $c_\P$ and $\tilde c_P$ are constants that can be fit for numerically, and describe the non-perturbative and perturbative scalings respectively. We note that although it may appear unnatural, the coefficients $c_\NP$ and $\tilde c_P$ have different mass dimensions, since $c_\NP$ is associated with a power-law variation, while $\tilde c_P$ is associated with a logarithmic variation.

From a practical perspective, the CSS decorrelation function can be constructed by fixing the value of $\alpha$ appearing in \Eq{eq:shapefunc} using a single value of the mass. For $D_2$, we find values of $\alpha \in [2,3]$ work well, with no strong preference for a given value. Using several values of the mass, one can then fit for $c_\NP$ and $\tilde c_P$ to give a smooth function that describes the evolution of the moment of the shape function. Knowing the analytic scaling of the function is therefore important, as it allows the shape to be fixed using dedicated Monte Carlo at a few specific mass points, and does not require Monte Carlo at every single value of the mass to determine the form. We will illustrate this for a case study of $Z'\to q\bar q$ in \Sec{sec:ph}, where we will find that this gives a remarkably good (almost perfect) decorrelation of the $D_2$ observable.

\subsection{Practical Implementation}\label{sec:practical}

\begin{figure}
\begin{center}
\subfloat[]{\label{fig:cssconstruction_a}
\includegraphics[width=7cm]{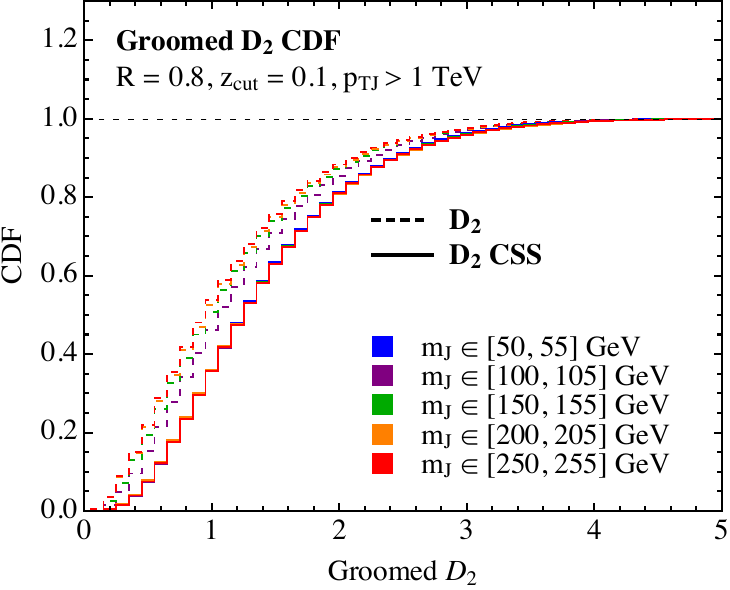}    
}\qquad
\subfloat[]{\label{fig:cssconstruction_b}
\includegraphics[width=6.8cm]{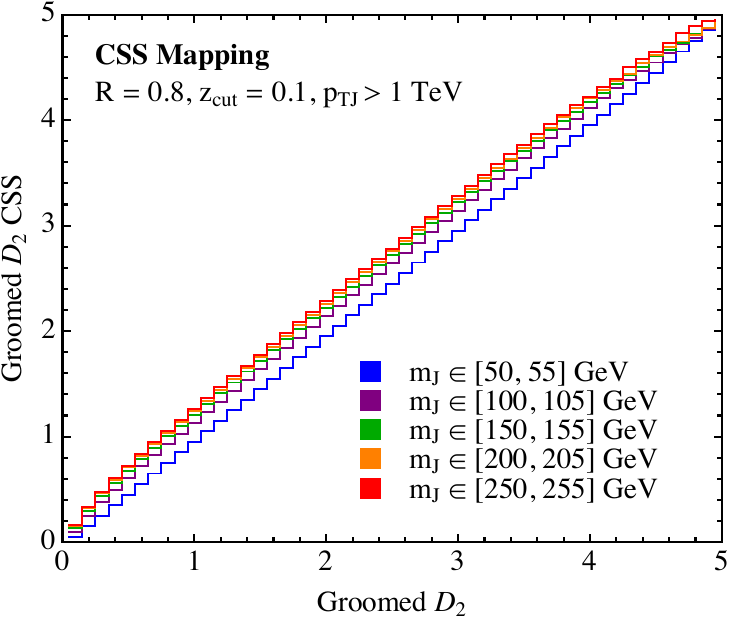}
}
\end{center}
\caption{(a) The CDF of the $D_2$ and $D_2$ CSS distributions in various bins of groomed mass for QCD jets. (b) The mapping between $D_2$  and $D_2$ CSS. Here an angular exponent $\beta=2$ was used for $D_2$. The mapping is linear throughout most of the range of interest, but with important non-linearities at small and large values of $D_2$. 
}
\label{fig:cssconstruction}
\end{figure}

In practice, the convolution procedure described above needs to be applied jet-by-jet and not at the distribution level.  The convolution of two distributions corresponds to the addition of the random variables described by the distributions.  Therefore, one possibility for translating the distribution-level results from earlier to event-by-event results is to add to every observed $D_2$ value a random value drawn from the distribution $F_\CSS(x;\alpha,\Omega_D)$ from Eq.~\ref{eq:shapefunc}.  This is not ideal because (a) the randomness can introduce features in the classification performance for finite statistics and (b) there are various technical reasons like reproducibility that make injecting randomness unattractive.  Another way to accomplish the convolution but using a deterministic approach is to use the (inverse) cumulative distribution function (CDF).  Given a random variable $X$ with CDF $C(x)=\Pr(X<x)$, $C(X)$ is a new random variable that follows a uniform distribution.  For any other CDF $G$, one can then form the random variable $G^{-1}(C(X))$, which follows the probability distribution $g(x)=\partial_y G(y)|_{y=x}$ that corresponds to $G$.  Let
\begin{align}
c(x)&=\frac{1}{\sigma}\frac{d\sigma}{dD_2}\\
g(x;\alpha,\Delta_D)&=c(x)\otimes F_\CSS(x;\alpha,\Delta_D).
\end{align}
We can now define the CDFs $C(x)=\int_{0}^x c(x')dx'$ and $G(x;\alpha,\Delta_D)=\int_{0}^x g(x';\alpha,\Delta_D)dx'$.  Then, the jet-by-jet transformation is given by
\begin{align}
\label{eq:mapping}
D_2\mapsto G^{-1}(C(D_2);\alpha,\Omega_D).
\end{align}
This simple mapping allows us to numerically implement the CSS procedure in an efficient manner.

An explicit example of the mapping given by Eq.~\ref{eq:mapping} for the example of $Z'\to q \bar q$, which is discussed in detail in \Sec{sec:ph}, is shown in \Fig{fig:cssconstruction}. This figure demonstrates the construction of the CSS $D_2$, following the procedure from Sec.~\ref{sec:practical}.  The CDF for each $D_2$ distribution is computed ($C$ for $D_2$ and $G$ for $ D_2$ CSS), as shown in \Fig{fig:cssconstruction_a} and then the transformation in Eq.~\ref{eq:mapping} is shown in \Fig{fig:cssconstruction_b}.  While the CSS curves may look mostly linear, there are important non-linear features at high and low $D_2$. These will be discussed in detail in \Sec{sec:ph}, and will play an important role in decorrelating the complete $D_2$ distribution, and not just the first moment. The perturbative expansion of the CSS procedure to its first moment, as was discussed around \Eq{eq:taylor_CSS} gives rise to a linear behavior, and the fact that the mapping in \Fig{fig:cssconstruction_b} is mostly linear simply shows that this is a good approximation.  Note that the DDT procedure would result in straight lines in Fig.~\ref{fig:cssconstruction} with a mass-dependent offset.

\begin{figure}
\begin{center}
\subfloat[]{\label{fig:D2}
\includegraphics[width=7cm]{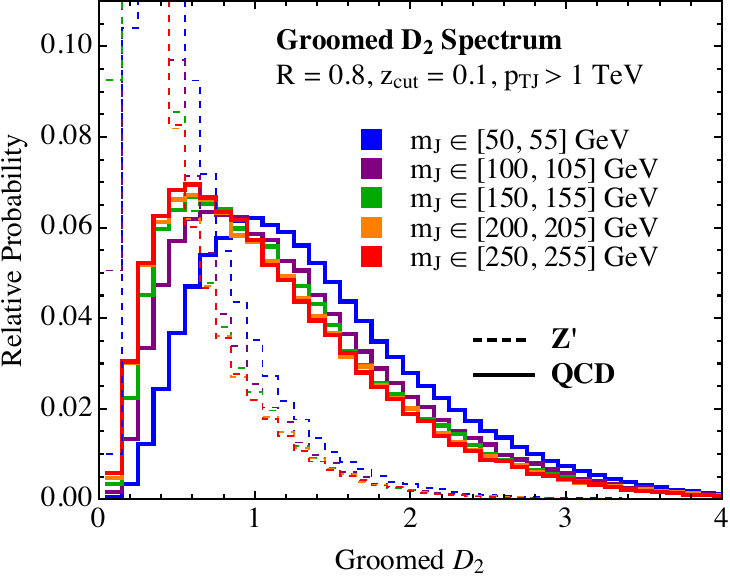}    
}\qquad
\subfloat[]{\label{fig:D2_CSS}
\includegraphics[width=7cm]{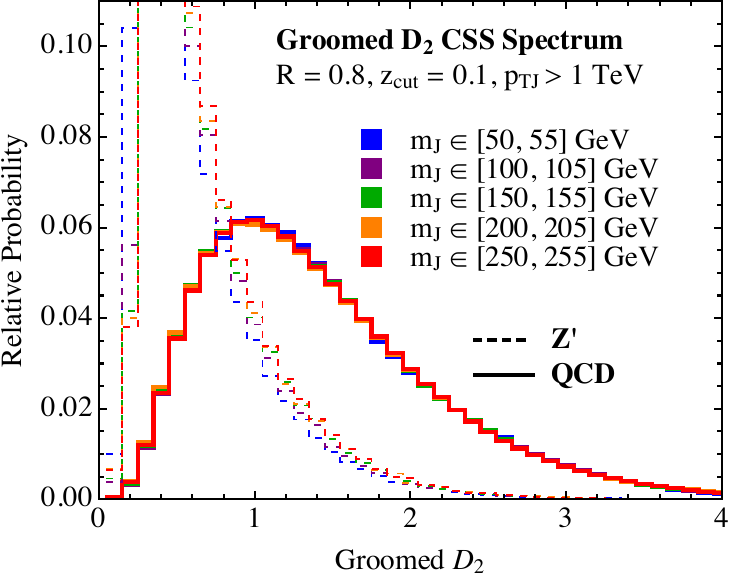}
}\\
\subfloat[]{\label{fig:D2_ddt}
\includegraphics[width=7cm]{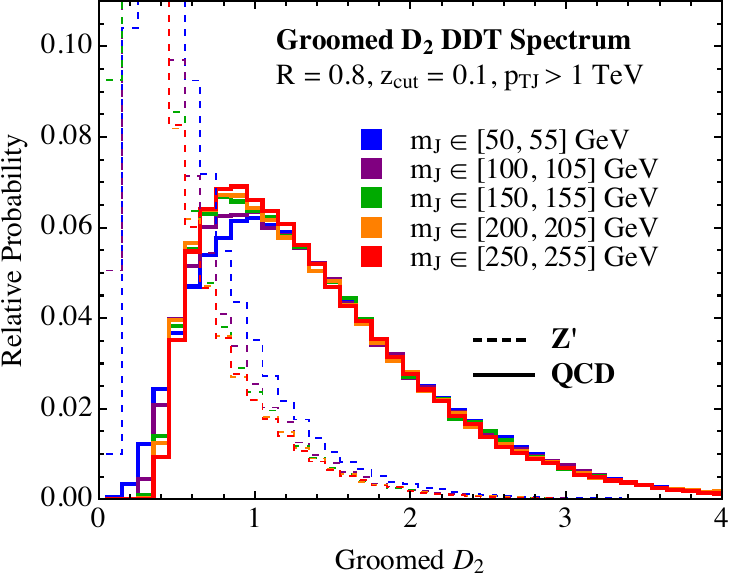}    
}\qquad
\subfloat[]{\label{fig:D2_diff}
\includegraphics[width=7cm]{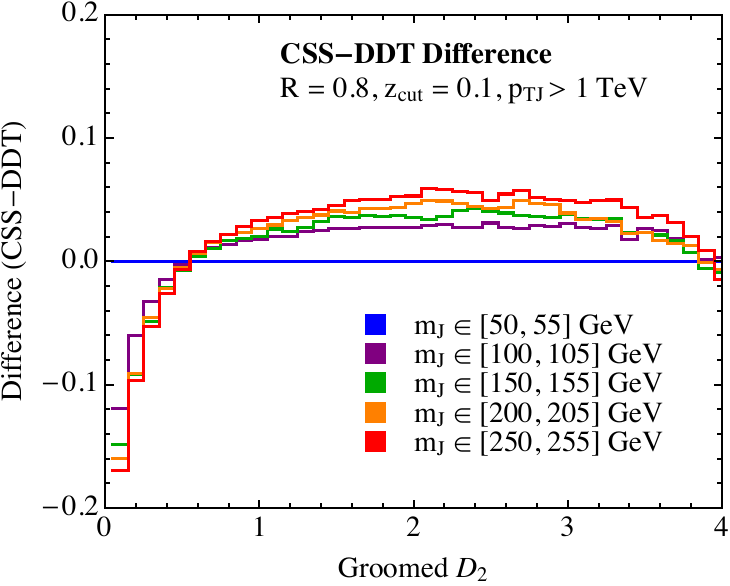}
}
\end{center}
\caption{Figure (a) shows the groomed $D_2$ distribution without further modification. Figures (b) and (c) show the groomed $D_2$ distribution after the application of the CSS and DDT procedures, respectively. The differences between the DDT and CSS distributions are shown in Figure (d), and grow at small values.  The DDT and CSS procedures are applied to both signal and background, where the transformations are defined by the background distributions.
}
\label{fig:D2_func}
\end{figure}

\section{A Case Study: $D_2$ for $Z'\to q\bar q$}\label{sec:ph}

\begin{figure}
\begin{center}
\subfloat[]{\label{fig:D2_DDT_zoom}
\includegraphics[width=7cm]{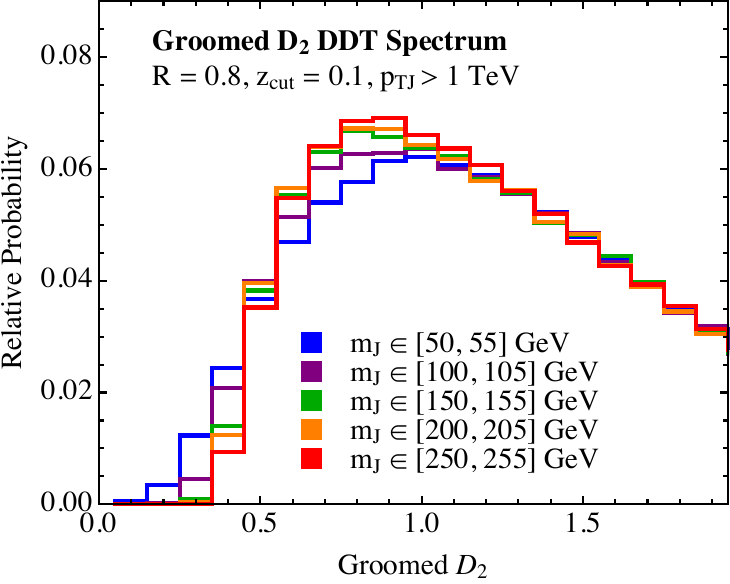}    
}\qquad
\subfloat[]{\label{fig:D2_CSS_zoom}
\includegraphics[width=7cm]{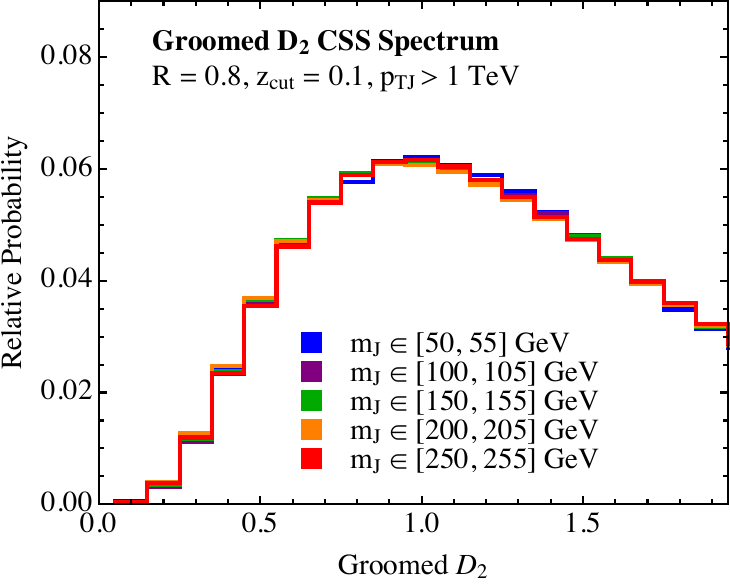}
}
\end{center}
\caption{A comparison of the groomed $D_2$ DDT distribution in (a) and the groomed $D_2$ CSS distribution in (b) at small values. The CSS approach decorrelates the entire shape of the distribution, including at low values of $D_2$, where the shape of the distribution changes non-trivially, and which is the relevant region for discrimination. 
}
\label{fig:D2_func_zoom}
\end{figure}

An important and recent application of variable decorrelation is the search for a low mass hadronic resonance, $Z'\rightarrow q\bar{q}$~\cite{CMS-PAS-EXO-17-001,Sirunyan:2017dnz,Sirunyan:2017nvi}, which we therefore use as a case study.  The generic quark and gluon background is too large to observe a dijet resonance directly, but when the $Z'$ is produced in association with initial state radiation, it can be sufficiently boosted for its decay products to be collimated inside a single jet.  For our study, both the $Z'$ and the generic quark and gluon background are simulated with \pythia{8.183} \cite{Sjostrand:2006za,Sjostrand:2014zea}; the former by changing the mass of a Standard Model $Z$ boson and the latter with all hard QCD processes.  All stable final state particles excluding neutrinos and muons are clustered into jets with \fastjet{3.1.3} \cite{Cacciari:2011ma} using the anti-$k_t$ algorithm~\cite{Cacciari:2008gp,Cacciari:2011ma} with $R = 0.8$.  In order to make sure that the $Z'$ particles with masses up to $300$ GeV are mostly contained inside a single jet, jets are required to have $p_\text{T}>1$ TeV.  Jets are then re-clustered using the Cambridge/Aachen algorithm \cite{Dokshitzer:1997in,Wobisch:1998wt,Wobisch:2000dk} and groomed with mMDT/soft drop using $z_\text{cut}=0.1$.  From the groomed jet's constituents, the jet mass is calculated along with $D_2$ using the \texttt{EnergyCorrelator} \fastjet{contrib} \cite{Cacciari:2011ma,fjcontrib}.  Throughout this section we will use an angular exponent of $\beta=2$ for the $D_2$ observable, but for notational simplicity, we will suppress the argument.

To perform the CSS decorrelation, we will shift all distributions to the reference mass of $m=50$ GeV, and we will consider jets with masses in the range $50$ GeV $<m<$ $250$ GeV, namely a factor of $5$ variation. This is approximately the mass range used in the current LHC searches \cite{CMS-PAS-EXO-17-001,Sirunyan:2017dnz,Sirunyan:2017nvi}. In a realistic application, it may be convenient to shift the distributions in different mass regions to different reference values. For example, for low mass searches, the $Z$ mass provides a natural mass scale where the analysis changes, and therefore it may prove useful to shift jets with mass $m>m_Z$ to the reference mass of $m_Z$, and jets with mass $m< m_Z$ to the lower mass limit of the search. In this way, the required decorrelation in each mass window is minimized. However, the goal of this section is simply to illustrate that we can completely decorrelate the $D_2$ distribution over a wide range of jet masses.

In \Fig{fig:D2_func} we show the standard $D_2$ distribution, as well as the decorrelated distributions using the CSS and DDT approaches, for five narrow bins in the groomed jet mass.  The DDT is applied by shifting 

\begin{align}
D_2\mapsto D_2-\langle D_2|m\rangle+\langle D_2|50 < m/\text{GeV}<55\rangle,
\end{align}

\noindent where the averages $\langle x|y\rangle$ (this means the average of $x$ given $y$) are computed using the QCD background jets.  By construction, the average of the resulting DDT distribution is independent of $m$:  

\begin{align}
\langle D_2 \text{ DDT}| m \rangle &= \langle D_2-\langle D_2|m\rangle+\langle D_2|50 < m/\text{GeV}<55\rangle | m\rangle\\
&= \langle D_2|m\rangle-\langle D_2|m\rangle+\langle D_2|50 < m/\text{GeV}<55\rangle\\
&=\langle D_2|50 < m/\text{GeV}<55\rangle
\end{align}

\begin{figure}
\begin{center}
\includegraphics[width=8cm]{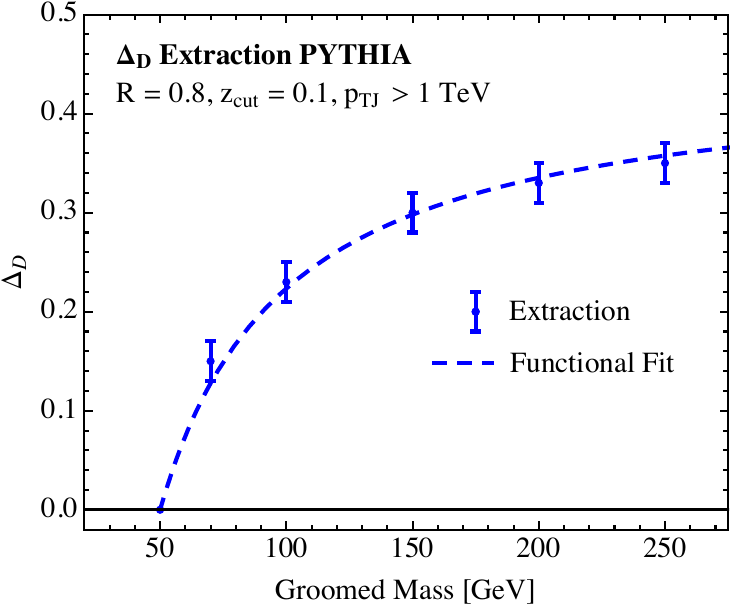}    
\end{center}
\caption{The first moment of the CSS mapping, $\Delta_D$, as a function of jet mass as extracted from \pythia{}, and compared with a fit to the analytic form described in \Eq{eq:total_moment_formula} in the text.
}
\label{fig:final_shift}
\end{figure}

\noindent The CSS procedure is applied using the shape function, $F_\CSS$, of Eq.~\ref{eq:shapefunc} with $\alpha=2.4$ and $\Omega_D$ as indicated in the figure. The value of $\alpha$ was fixed for a single value of the mass, however, fortunately, we find that we are quite insensitive to the precise choice of $\alpha$.  The values of $\Omega_D$ are plotted in \Fig{fig:final_shift} along with a fit to the analytic form, which we see provides an excellent description. The extractions of the shift at these five mass values can be viewed as fixing the coefficients of the analytic mass dependence of the decorrelation procedure of \Eq{eq:total_moment_formula}, and providing a prediction for every other value of the mass, as would be required experimentally. Here we see the advantage of knowing the analytic form, namely that one only needs dedicated Monte Carlo at several specific mass values. The signal distributions are also shown to give a feeling for the range of interest of the $D_2$ observable for discrimination.

A number of features of the different decorrelation procedures are clearly evident from these figures. First, we see in \Fig{fig:D2_CSS} that the CSS decorrelated observable has essentially no dependence on the jet mass. The complete shape of the distribution is identical for the wide range of masses shown. By contrast, the shape of the DDT version in  \Fig{fig:D2_ddt} changes with mass even though the mean is fixed.  This is particularly true on the left side of the peak. The difference between methods arises from the non-linear nature of the CSS mapping, as was mentioned in \Fig{fig:cssconstruction}. A zoomed in view of the small $D_2$ region is shown in \Fig{fig:D2_func_zoom}, which highlights the difference between the two approaches. Both the CSS and DDT mappings are effectively linear to the right of the $D_2$ peak, where we see that both decorrelate the observable very well, but the non-linear mapping is required to perform the decorrelation of the shape of the distribution at small values of $D_2$. It is in this region that the shape of the distribution changes non-trivially with mass, and the difference between distributions at different masses cannot simply be described by a shift. The ability of the CSS approach to correctly reproduce the change in shape of the distribution in this region of the distribution, which is the most important region for discrimination, is quite remarkable. The differences between the two different decorrelated distributions are shown in \Fig{fig:D2_diff}, which also highlights that the differences between the two decorrelation procedures become large at small values of $D_2$. We also note that here we have chosen to decorrelate the background (QCD) distributions, and therefore the signal distributions exhibit some dependence on mass.

\begin{figure}
\begin{center}
\subfloat[]{\label{fig:means_a}
\includegraphics[width=7cm]{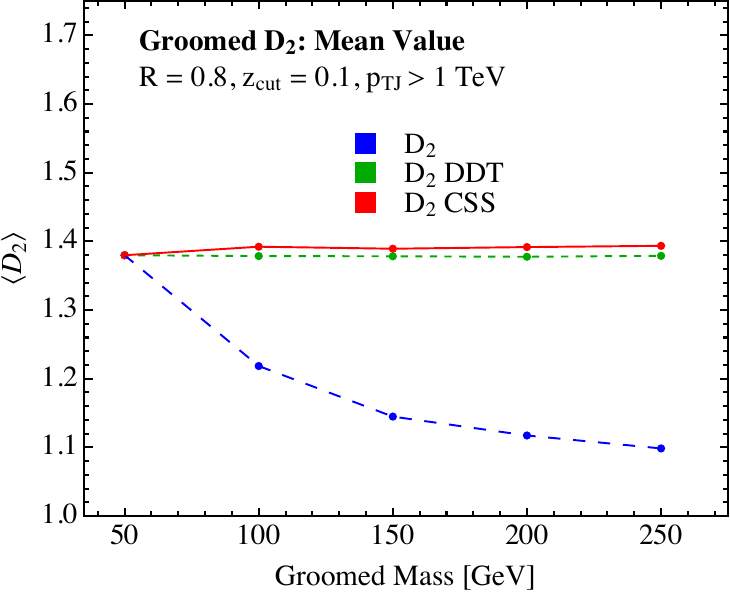}    
}\qquad
\subfloat[]{\label{fig:means_b}
\includegraphics[width=7.1cm]{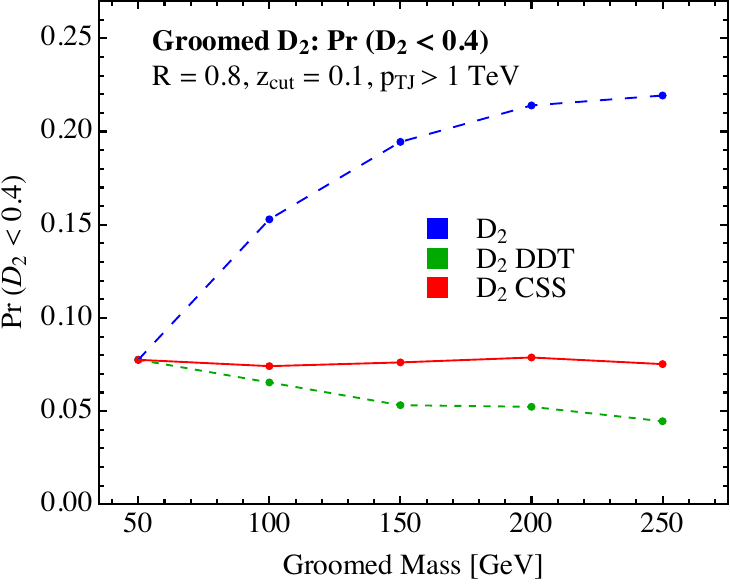}
}
\end{center}
\caption{(a) The mean $D_2$ and (b) $\Pr(D_2 < 0.4)$  for QCD jets as a function of mass. By construction the DDT decorrelates a single moment, chosen here to be the first moment, but does not decorrelate higher moments. On the other hand, the CSS procedure is designed to decorrelate the entire shape of the distribution.
}
\label{fig:means}
\end{figure}

As a further quantitative comparison between the CSS and DDT approaches, in \Fig{fig:means} we compare different integrals of the distributions, namely the mean, and the probability that $D_2 \leq0.4$ (the lower tail fraction), which we denote $\Pr(D_2 < 0.4)$. By construction, the mean of the DDT $D_2$ distribution is independent of mass, as seen in \Fig{fig:means_a}.  However, the shape does change with mass as indicated by the lower tail fraction in \Fig{fig:means_b} (lower $D_2$ is more signal-like). On the other hand, since the CSS approach decorrelates the complete shape of the distribution both the mean and the tail fraction are nearly independent of mass. We must also emphasize that the DDT approach could equally well be applied to flatten the $\Pr(D_2^{(2)} < 0.4)$ (or any other given integral of the distribution). However, it would then not decorrelate the mean. In other words, it can be used to decorrelate a single moment at a time. On the other hand, the CSS approach aims to decorrelate all moments.

\begin{figure}
\begin{center}
\subfloat[]{\label{fig:roc_a}
\includegraphics[width=0.48\textwidth]{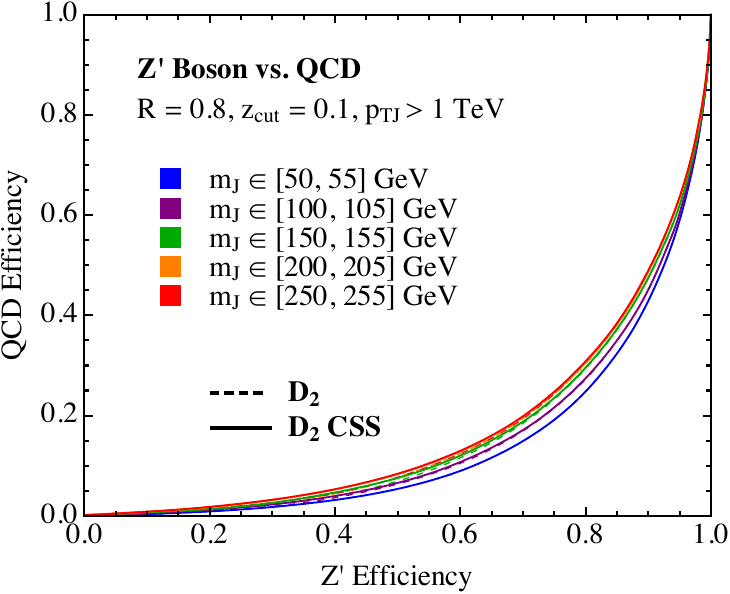}   }
 \subfloat[]{\label{fig:roc_b}
\includegraphics[width=0.5\textwidth]{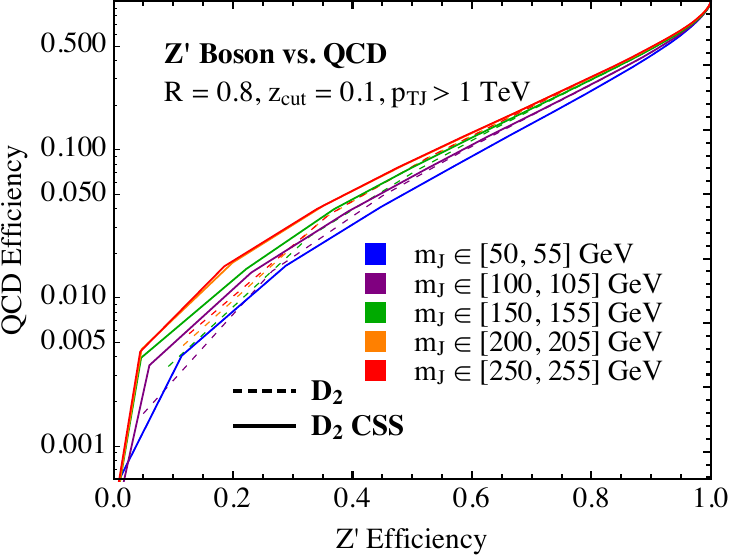}   }
\end{center}
\caption{A scan in an upper cut on $D_2$ traces out a Receiver Operator Characteristic (ROC) curve quantifying the tradeoff between $Z$' (signal) efficiency and QCD (background) efficiency for various groomed jet mass bins, shown in a linear plot in a) and a log plot in b). The CSS procedure is not found to significantly degrade the discrimination power of the observable.}
\label{fig:roc}
\end{figure}

Finally, it is important to check that the CSS procedure does not degrade the tagging performance of the $D_2$ observable. This was shown for the DDT approach in \cite{Dolen:2016kst}.
Applying the mapping shown in the right plot of \Fig{fig:cssconstruction} also to $Z'$ events results in the distributions that were already shown in \Fig{fig:D2_func}.  Lower values of $D_2$ are more signal-like so an upper-threshold on the $D_2$ distribution is an effective two-prong tagger.  \Fig{fig:roc} quantifies the tradeoff between signal and background efficiency with and without the CSS procedure.  As desired, there is a minimal difference in the ROC curve after applying CSS. This difference could be further minimized by performing the CSS decorrelation in narrower mass windows.

\section{Conclusions}\label{sec:conc}

In this paper, we have shown how a given jet substructure observable, such as $N_2$ or $D_2$, can be decorrelated with the jet mass using an understanding of its perturbative and non-perturbative behavior. Inspired by the use of shape functions for modeling non-perturbative effects, we introduced the Convolved SubStructure (CSS) approach, which uses a shape function, convolved with the substructure observable's distribution, to map it to a reference mass. The shape function incorporates effects due to both perturbative and non-perturbative physics, and we used a recently derived factorization formula to analytically derive the mass dependence of both these contributions.  Unlike previous approaches with similar philosophies, the CSS approach completely decorrelates the entire shape of the distribution. Furthermore, it is systematically improvable by expanding the shape function in a basis of orthogonal functions \cite{Ligeti:2008ac}, and uses maximally the theoretical understanding of the observable.  

We have shown in detail how the CSS approach can be practically implemented in an extremely simple manner, and studied its behavior for the example of a light $Z'\to q\bar q$ search using the $D_2$ observable. We found that using a simple two parameter shape function we were able to obtain an excellent decorrelation of the entire $D_2$ distribution over a wide range of mass values. The shape function parameter defining the shift of the first moment of the distribution has a functional dependence on the jet mass that can be understood from first principles, and is fixed by demanding that the first moment of the mapped distributions are the same as the reference mass distribution. Higher moments can be handled similarly, but since we require the shape function to maintain the domain and norm of the distribution, we find that already the decorrelation of the first moment effectively decorrelates the whole spectrum. Furthermore, the discrimination power of the CSS observable was not significantly degraded. In real applications, the tradeoff between discrimination power and decorrelation must be evaluated, and it may be practical to perform the decorrelation in mass windows.

One important aspect that we did not study in this paper is whether an identical mapping applies at detector level. Even if it is not the case, our approach is general, and another simple functional form that performs the decorrelation could be found. It will also be interesting to apply the CSS approach to other observables, such as $N_2$, for which the DDT approach has been applied successfully \cite{CMS-PAS-EXO-17-001,Sirunyan:2017nvi}. Again, a slightly modified convolution function may be required, depending on the behavior of the observable. We therefore hope that the CSS approach can be used to decorrelate a variety of substructure observables, improving the reach and performance of searches for low mass particles at the LHC.

\begin{acknowledgments}
We thank Nhan Tran and Simone Marzani for helpful and encouraging discussions, and Andrew Larkoski for collaboration on related topics and for catching several typos in the first draft of this work. This work is supported by the U.S. Department of Energy (DOE) under cooperative research agreements DE-FG02-05ER-41360, and DE-SC0011090 and by the Office of High Energy Physics of the U.S. DOE under Contract No. DE-AC02-05CH11231, and the LDRD Program of LBNL. This work was performed in part at the Aspen Center for Physics, which is supported by National Science Foundation grant PHY-1607611. D.N. thanks both the Berkeley Center for Theoretical Physics and LBNL for hospitality while portions of this work were completed, as well as support from the Munich Institute for Astro- and Particle Physics (MIAPP) of the DFG cluster of excellence “Origin and Structure of the Universe”, and support from DOE contract DE-AC52-06NA25396 at LANL and through the LANL/LDRD Program.
\end{acknowledgments}

\bibliography{ddt_bib}

\providecommand{\href}[2]{#2}\begingroup\raggedright\begin{thebibliography}{10}

\bibitem{ATLAS-CONF-2015-035}
{\scshape ATLAS} collaboration, \emph{{Performance of jet substructure
  techniques in early $\sqrt{s}=13$ TeV $pp$ collisions with the ATLAS
  detector}}, {\emph{ATLAS-CONF-2015-035} (2015) }.

\bibitem{Aad:2015rpa}
{\scshape ATLAS} collaboration, G.~Aad et~al., \emph{{Identification of
  Boosted, Hadronically Decaying W Bosons and Comparisons with ATLAS Data Taken
  at $\sqrt{s} = 8$ TeV}},  \href{http://arXiv.org/abs/1510.05821}{{\tt
  arXiv:1510.05821}}.

\bibitem{Aaboud:2016okv}
{\scshape ATLAS} collaboration, M.~Aaboud et~al., \emph{{Searches for heavy
  diboson resonances in $pp$ collisions at $\sqrt{s}=13$ TeV with the ATLAS
  detector}}, \href{http://dx.doi.org/10.1007/JHEP09(2016)173}{\emph{JHEP} {\bf
  09} (2016) 173}, [\href{http://arXiv.org/abs/1606.04833}{{\tt
  arXiv:1606.04833}}].

\bibitem{Aaboud:2016trl}
{\scshape ATLAS} collaboration, M.~Aaboud et~al., \emph{{Search for heavy
  resonances decaying to a $Z$ boson and a photon in $pp$ collisions at
  $\sqrt{s}=13$ TeV with the ATLAS detector}},
  \href{http://dx.doi.org/10.1016/j.physletb.2016.11.005}{\emph{Phys. Lett.}
  {\bf B764} (2017) 11--30}, [\href{http://arXiv.org/abs/1607.06363}{{\tt
  arXiv:1607.06363}}].

\bibitem{Aaboud:2016qgg}
{\scshape ATLAS} collaboration, M.~Aaboud et~al., \emph{{Search for dark matter
  produced in association with a hadronically decaying vector boson in $pp$
  collisions at $\sqrt{s} =$ 13 TeV with the ATLAS detector}},
  \href{http://dx.doi.org/10.1016/j.physletb.2016.10.042}{\emph{Phys. Lett.}
  {\bf B763} (2016) 251--268}, [\href{http://arXiv.org/abs/1608.02372}{{\tt
  arXiv:1608.02372}}].

\bibitem{Aaboud:2017zfn}
{\scshape ATLAS} collaboration, M.~Aaboud et~al., \emph{{Search for pair
  production of heavy vector-like quarks decaying to high-$p_{\mathrm{T}}$ $W$
  bosons and $b$ quarks in the lepton-plus-jets final state in $pp$ collisions
  at $\sqrt{s}$=13 TeV with the ATLAS detector}},
  \href{http://arXiv.org/abs/1707.03347}{{\tt arXiv:1707.03347}}.

\bibitem{Aaboud:2017ahz}
{\scshape ATLAS} collaboration, M.~Aaboud et~al., \emph{{Search for heavy
  resonances decaying to a $W$ or $Z$ boson and a Higgs boson in the
  $q\bar{q}^{(\prime)}b\bar{b}$ final state in $pp$ collisions at $\sqrt{s} =
  13$ TeV with the ATLAS detector}},
  \href{http://arXiv.org/abs/1707.06958}{{\tt arXiv:1707.06958}}.

\bibitem{Aaboud:2017eta}
{\scshape ATLAS} collaboration, M.~Aaboud et~al., \emph{{Search for diboson
  resonances with boson-tagged jets in $pp$ collisions at $\sqrt{s}=13$ TeV
  with the ATLAS detector}},  \href{http://arXiv.org/abs/1708.04445}{{\tt
  arXiv:1708.04445}}.

\bibitem{Aaboud:2017itg}
{\scshape ATLAS} collaboration, M.~Aaboud et~al., \emph{{Searches for heavy
  $ZZ$ and $ZW$ resonances in the $\ell\ell qq$ and $\nu\nu qq$ final states in
  $pp$ collisions at $\sqrt{s}=13$ TeV with the ATLAS detector}},
  \href{http://arXiv.org/abs/1708.09638}{{\tt arXiv:1708.09638}}.

\bibitem{Aaboud:2017ecz}
{\scshape ATLAS} collaboration, M.~Aaboud et~al., \emph{{A search for
  resonances decaying into a Higgs boson and a new particle $X$ in the $XH \to
  qqbb$ final state with the ATLAS detector}},
  \href{http://arXiv.org/abs/1709.06783}{{\tt arXiv:1709.06783}}.

\bibitem{Khachatryan:2015bma}
{\scshape CMS} collaboration, V.~Khachatryan et~al., \emph{{Search for a
  massive resonance decaying into a Higgs boson and a W or Z boson in hadronic
  final states in proton-proton collisions at sqrt(s) = 8 TeV}},
  \href{http://arXiv.org/abs/1506.01443}{{\tt arXiv:1506.01443}}.

\bibitem{Khachatryan:2016cfa}
{\scshape CMS} collaboration, V.~Khachatryan et~al., \emph{{Search for heavy
  resonances decaying to two Higgs bosons in final states containing four b
  quarks}}, \href{http://dx.doi.org/10.1140/epjc/s10052-016-4206-6}{\emph{Eur.
  Phys. J.} {\bf C76} (2016) 371}, [\href{http://arXiv.org/abs/1602.08762}{{\tt
  arXiv:1602.08762}}].

\bibitem{Sirunyan:2016cao}
{\scshape CMS} collaboration, A.~M. Sirunyan et~al., \emph{{Search for massive
  resonances decaying into WW, WZ or ZZ bosons in proton-proton collisions at
  $\sqrt{s} = $ 13 TeV}},
  \href{http://dx.doi.org/10.1007/JHEP03(2017)162}{\emph{JHEP} {\bf 03} (2017)
  162}, [\href{http://arXiv.org/abs/1612.09159}{{\tt arXiv:1612.09159}}].

\bibitem{Sirunyan:2017acf}
{\scshape CMS} collaboration, A.~M. Sirunyan et~al., \emph{{Search for massive
  resonances decaying into WW, WZ, ZZ, qW, and qZ with dijet final states at
  sqrt(s) = 13 TeV}},  \href{http://arXiv.org/abs/1708.05379}{{\tt
  arXiv:1708.05379}}.

\bibitem{Sirunyan:2017wto}
{\scshape CMS} collaboration, A.~M. Sirunyan et~al., \emph{{Search for heavy
  resonances that decay into a vector boson and a Higgs boson in hadronic final
  states at $\sqrt{s} = 13$ $\,\text {TeV}$}},
  \href{http://dx.doi.org/10.1140/epjc/s10052-017-5192-z}{\emph{Eur. Phys. J.}
  {\bf C77} (2017) 636}, [\href{http://arXiv.org/abs/1707.01303}{{\tt
  arXiv:1707.01303}}].

\bibitem{Sirunyan:2017ukk}
{\scshape CMS} collaboration, A.~M. Sirunyan et~al., \emph{{Searches for W′
  bosons decaying to a top quark and a bottom quark in proton-proton collisions
  at 13 TeV}}, \href{http://dx.doi.org/10.1007/JHEP08(2017)029}{\emph{JHEP}
  {\bf 08} (2017) 029}, [\href{http://arXiv.org/abs/1706.04260}{{\tt
  arXiv:1706.04260}}].

\bibitem{Sirunyan:2017bfa}
{\scshape CMS} collaboration, A.~M. Sirunyan et~al., \emph{{Search for a heavy
  resonance decaying to a top quark and a vector-like top quark at $
  \sqrt{s}=13 $ TeV}},
  \href{http://dx.doi.org/10.1007/JHEP09(2017)053}{\emph{JHEP} {\bf 09} (2017)
  053}, [\href{http://arXiv.org/abs/1703.06352}{{\tt arXiv:1703.06352}}].

\bibitem{Sirunyan:2017uhk}
{\scshape CMS} collaboration, A.~M. Sirunyan et~al., \emph{{Search for $
  \mathrm{t}\overline{\mathrm{t}} $ resonances in highly boosted lepton+jets
  and fully hadronic final states in proton-proton collisions at $ \sqrt{s}=13
  $ TeV}}, \href{http://dx.doi.org/10.1007/JHEP07(2017)001}{\emph{JHEP} {\bf
  07} (2017) 001}, [\href{http://arXiv.org/abs/1704.03366}{{\tt
  arXiv:1704.03366}}].

\bibitem{Sirunyan:2017dgc}
{\scshape CMS} collaboration, A.~M. Sirunyan et~al., \emph{{Inclusive search
  for a highly boosted Higgs boson decaying to a bottom quark-antiquark pair}},
   \href{http://arXiv.org/abs/1709.05543}{{\tt arXiv:1709.05543}}.

\bibitem{CMS-PAS-HIG-17-010}
{\scshape CMS Collaboration} collaboration, \emph{{Inclusive search for the
  standard model Higgs boson produced in pp collisions at
  $\sqrt{s}=13~\mathrm{TeV}$ using H$\rightarrow \mathrm{b\bar{\mathrm{b}}}$
  decays}},  Tech. Rep. CMS-PAS-HIG-17-010, CERN, Geneva, 2017.

\bibitem{CMS-PAS-EXO-17-001}
{\scshape CMS Collaboration} collaboration, \emph{{Search for light vector
  resonances decaying to a quark pair produced in association with a jet in
  proton-proton collisions at $\sqrt{s}=13~\mathrm{TeV}$}},  Tech. Rep.
  CMS-PAS-EXO-17-001, CERN, Geneva, 2017.

\bibitem{Sirunyan:2017dnz}
{\scshape CMS} collaboration, A.~M. Sirunyan et~al., \emph{{Search for Low Mass
  Vector Resonances Decaying to Quark-Antiquark Pairs in Proton-Proton
  Collisions at $\sqrt{s}=13\text{ }\text{ }\mathrm{TeV}$}},
  \href{http://dx.doi.org/10.1103/PhysRevLett.119.111802}{\emph{Phys. Rev.
  Lett.} {\bf 119} (2017) 111802}, [\href{http://arXiv.org/abs/1705.10532}{{\tt
  arXiv:1705.10532}}].

\bibitem{Sirunyan:2017nvi}
{\scshape CMS} collaboration, A.~M. Sirunyan et~al., \emph{{Search for low mass
  vector resonances decaying into quark-antiquark pairs in proton-proton
  collisions at $\sqrt{s} = $ 13 TeV}},
  \href{http://arXiv.org/abs/1710.00159}{{\tt arXiv:1710.00159}}.

\bibitem{Mariotti:2017vtv}
A.~Mariotti, D.~Redigolo, F.~Sala and K.~Tobioka, \emph{{New LHC bound on
  low-mass diphoton resonances}},  \href{http://arXiv.org/abs/1710.01743}{{\tt
  arXiv:1710.01743}}.

\bibitem{Dolen:2016kst}
J.~Dolen, P.~Harris, S.~Marzani, S.~Rappoccio and N.~Tran, \emph{{Thinking
  outside the ROCs: Designing Decorrelated Taggers (DDT) for jet
  substructure}}, \href{http://dx.doi.org/10.1007/JHEP05(2016)156}{\emph{JHEP}
  {\bf 05} (2016) 156}, [\href{http://arXiv.org/abs/1603.00027}{{\tt
  arXiv:1603.00027}}].

\bibitem{Shimmin:2017mfk}
C.~Shimmin, P.~Sadowski, P.~Baldi, E.~Weik, D.~Whiteson, E.~Goul et~al.,
  \emph{{Decorrelated Jet Substructure Tagging using Adversarial Neural
  Networks}},  \href{http://arXiv.org/abs/1703.03507}{{\tt arXiv:1703.03507}}.

\bibitem{Aguilar-Saavedra:2017rzt}
J.~A. Aguilar-Saavedra, J.~H. Collins and R.~K. Mishra, \emph{{A generic
  anti-QCD jet tagger}},  \href{http://arXiv.org/abs/1709.01087}{{\tt
  arXiv:1709.01087}}.

\bibitem{Larkoski:2017jix}
A.~J. Larkoski, I.~Moult and B.~Nachman, \emph{{Jet Substructure at the Large
  Hadron Collider: A Review of Recent Advances in Theory and Machine
  Learning}},  \href{http://arXiv.org/abs/1709.04464}{{\tt arXiv:1709.04464}}.

\bibitem{Marzani:2017mva}
S.~Marzani, L.~Schunk and G.~Soyez, \emph{{A study of jet mass distributions
  with grooming}},  \href{http://arXiv.org/abs/1704.02210}{{\tt
  arXiv:1704.02210}}.

\bibitem{Frye:2016aiz}
C.~Frye, A.~J. Larkoski, M.~D. Schwartz and K.~Yan, \emph{{Factorization for
  groomed jet substructure beyond the next-to-leading logarithm}},
  \href{http://dx.doi.org/10.1007/JHEP07(2016)064}{\emph{JHEP} {\bf 07} (2016)
  064}, [\href{http://arXiv.org/abs/1603.09338}{{\tt arXiv:1603.09338}}].

\bibitem{Frye:2016okc}
C.~Frye, A.~J. Larkoski, M.~D. Schwartz and K.~Yan, \emph{{Precision physics
  with pile-up insensitive observables}},
  \href{http://arXiv.org/abs/1603.06375}{{\tt arXiv:1603.06375}}.

\bibitem{Banfi:2016zlc}
A.~Banfi, H.~McAslan, P.~F. Monni and G.~Zanderighi, \emph{{The two-jet rate in
  $e^+e^-$ at next-to-next-to-leading-logarithmic order}},
  \href{http://dx.doi.org/10.1103/PhysRevLett.117.172001}{\emph{Phys. Rev.
  Lett.} {\bf 117} (2016) 172001}, [\href{http://arXiv.org/abs/1607.03111}{{\tt
  arXiv:1607.03111}}].

\bibitem{Banfi:2015pju}
A.~Banfi, F.~Caola, F.~A. Dreyer, P.~F. Monni, G.~P. Salam, G.~Zanderighi
  et~al., \emph{{Jet-vetoed Higgs cross section in gluon fusion at
  N$^{3}$LO+NNLL with small-$R$ resummation}},
  \href{http://dx.doi.org/10.1007/JHEP04(2016)049}{\emph{JHEP} {\bf 04} (2016)
  049}, [\href{http://arXiv.org/abs/1511.02886}{{\tt arXiv:1511.02886}}].

\bibitem{Stewart:2013faa}
I.~W. Stewart, F.~J. Tackmann, J.~R. Walsh and S.~Zuberi, \emph{{Jet $p_T$
  resummation in Higgs production at $NNLL'+NNLO$}},
  \href{http://dx.doi.org/10.1103/PhysRevD.89.054001}{\emph{Phys. Rev.} {\bf
  D89} (2014) 054001}, [\href{http://arXiv.org/abs/1307.1808}{{\tt
  arXiv:1307.1808}}].

\bibitem{Becher:2012qa}
T.~Becher and M.~Neubert, \emph{{Factorization and NNLL Resummation for Higgs
  Production with a Jet Veto}},
  \href{http://dx.doi.org/10.1007/JHEP07(2012)108}{\emph{JHEP} {\bf 07} (2012)
  108}, [\href{http://arXiv.org/abs/1205.3806}{{\tt arXiv:1205.3806}}].

\bibitem{Feige:2012vc}
I.~Feige, M.~D. Schwartz, I.~W. Stewart and J.~Thaler, \emph{{Precision Jet
  Substructure from Boosted Event Shapes}},
  \href{http://dx.doi.org/10.1103/PhysRevLett.109.092001}{\emph{Phys.Rev.Lett.}
  {\bf 109} (2012) 092001}, [\href{http://arXiv.org/abs/1204.3898}{{\tt
  arXiv:1204.3898}}].

\bibitem{Larkoski:2014wba}
A.~J. Larkoski, S.~Marzani, G.~Soyez and J.~Thaler, \emph{{Soft Drop}},
  \href{http://dx.doi.org/10.1007/JHEP05(2014)146}{\emph{JHEP} {\bf 1405}
  (2014) 146}, [\href{http://arXiv.org/abs/1402.2657}{{\tt arXiv:1402.2657}}].

\bibitem{Dasgupta:2013ihk}
M.~Dasgupta, A.~Fregoso, S.~Marzani and G.~P. Salam, \emph{{Towards an
  understanding of jet substructure}},
  \href{http://dx.doi.org/10.1007/JHEP09(2013)029}{\emph{JHEP} {\bf 1309}
  (2013) 029}, [\href{http://arXiv.org/abs/1307.0007}{{\tt arXiv:1307.0007}}].

\bibitem{Dasgupta:2013via}
M.~Dasgupta, A.~Fregoso, S.~Marzani and A.~Powling, \emph{{Jet substructure
  with analytical methods}},
  \href{http://dx.doi.org/10.1140/epjc/s10052-013-2623-3}{\emph{Eur.Phys.J.}
  {\bf C73} (2013) 2623}, [\href{http://arXiv.org/abs/1307.0013}{{\tt
  arXiv:1307.0013}}].

\bibitem{Dasgupta:2015lxh}
M.~Dasgupta, L.~Schunk and G.~Soyez, \emph{{Jet shapes for boosted jet
  two-prong decays from first-principles}},
  \href{http://dx.doi.org/10.1007/JHEP04(2016)166}{\emph{JHEP} {\bf 04} (2016)
  166}, [\href{http://arXiv.org/abs/1512.00516}{{\tt arXiv:1512.00516}}].

\bibitem{Larkoski:2015kga}
A.~J. Larkoski, I.~Moult and D.~Neill, \emph{{Analytic Boosted Boson
  Discrimination}},
  \href{http://dx.doi.org/10.1007/JHEP05(2016)117}{\emph{JHEP} {\bf 05} (2016)
  117}, [\href{http://arXiv.org/abs/1507.03018}{{\tt arXiv:1507.03018}}].

\bibitem{Frye:2017yrw}
C.~Frye, A.~J. Larkoski, J.~Thaler and K.~Zhou, \emph{{Casimir Meets Poisson:
  Improved Quark/Gluon Discrimination with Counting Observables}},
  \href{http://arXiv.org/abs/1704.06266}{{\tt arXiv:1704.06266}}.

\bibitem{Jouttenus:2013hs}
T.~T. Jouttenus, I.~W. Stewart, F.~J. Tackmann and W.~J. Waalewijn, \emph{{Jet
  mass spectra in Higgs boson plus one jet at next-to-next-to-leading
  logarithmic order}},
  \href{http://dx.doi.org/10.1103/PhysRevD.88.054031}{\emph{Phys.Rev.} {\bf
  D88} (2013) 054031}, [\href{http://arXiv.org/abs/1302.0846}{{\tt
  arXiv:1302.0846}}].

\bibitem{Dasgupta:2015yua}
M.~Dasgupta, A.~Powling and A.~Siodmok, \emph{{On jet substructure methods for
  signal jets}}, \href{http://dx.doi.org/10.1007/JHEP08(2015)079}{\emph{JHEP}
  {\bf 08} (2015) 079}, [\href{http://arXiv.org/abs/1503.01088}{{\tt
  arXiv:1503.01088}}].

\bibitem{Hoang:2017kmk}
A.~H. Hoang, S.~Mantry, A.~Pathak and I.~W. Stewart, \emph{{Extracting a Short
  Distance Top Mass with Light Grooming}},
  \href{http://arXiv.org/abs/1708.02586}{{\tt arXiv:1708.02586}}.

\bibitem{Larkoski:2017iuy}
A.~J. Larkoski, I.~Moult and D.~Neill, \emph{{Analytic Boosted Boson
  Discrimination at the Large Hadron Collider}},
  \href{http://arXiv.org/abs/1708.06760}{{\tt arXiv:1708.06760}}.

\bibitem{Larkoski:2017cqq}
A.~J. Larkoski, I.~Moult and D.~Neill, \emph{{Factorization and Resummation for
  Groomed Multi-Prong Jet Shapes}},
  \href{http://arXiv.org/abs/1710.00014}{{\tt arXiv:1710.00014}}.

\bibitem{Larkoski:2014gra}
A.~J. Larkoski, I.~Moult and D.~Neill, \emph{{Power Counting to Better Jet
  Observables}}, \href{http://dx.doi.org/10.1007/JHEP12(2014)009}{\emph{JHEP}
  {\bf 1412} (2014) 009}, [\href{http://arXiv.org/abs/1409.6298}{{\tt
  arXiv:1409.6298}}].

\bibitem{Bauer:2000yr}
C.~W. Bauer, S.~Fleming, D.~Pirjol and I.~W. Stewart, \emph{{An Effective field
  theory for collinear and soft gluons: Heavy to light decays}},
  \href{http://dx.doi.org/10.1103/PhysRevD.63.114020}{\emph{Phys.Rev.} {\bf
  D63} (2001) 114020}, [\href{http://arXiv.org/abs/hep-ph/0011336}{{\tt
  hep-ph/0011336}}].

\bibitem{Bauer:2001ct}
C.~W. Bauer and I.~W. Stewart, \emph{{Invariant operators in collinear
  effective theory}},
  \href{http://dx.doi.org/10.1016/S0370-2693(01)00902-9}{\emph{Phys.Lett.} {\bf
  B516} (2001) 134--142}, [\href{http://arXiv.org/abs/hep-ph/0107001}{{\tt
  hep-ph/0107001}}].

\bibitem{Bauer:2001yt}
C.~W. Bauer, D.~Pirjol and I.~W. Stewart, \emph{{Soft collinear factorization
  in effective field theory}},
  \href{http://dx.doi.org/10.1103/PhysRevD.65.054022}{\emph{Phys.Rev.} {\bf
  D65} (2002) 054022}, [\href{http://arXiv.org/abs/hep-ph/0109045}{{\tt
  hep-ph/0109045}}].

\bibitem{Bauer:2002nz}
C.~W. Bauer, S.~Fleming, D.~Pirjol, I.~Z. Rothstein and I.~W. Stewart,
  \emph{{Hard scattering factorization from effective field theory}},
  \href{http://dx.doi.org/10.1103/PhysRevD.66.014017}{\emph{Phys.Rev.} {\bf
  D66} (2002) 014017}, [\href{http://arXiv.org/abs/hep-ph/0202088}{{\tt
  hep-ph/0202088}}].

\bibitem{Rothstein:2016bsq}
I.~Z. Rothstein and I.~W. Stewart, \emph{{An Effective Field Theory for Forward
  Scattering and Factorization Violation}},
  \href{http://dx.doi.org/10.1007/JHEP08(2016)025}{\emph{JHEP} {\bf 08} (2016)
  025}, [\href{http://arXiv.org/abs/1601.04695}{{\tt arXiv:1601.04695}}].

\bibitem{Bauer:2011uc}
C.~W. Bauer, F.~J. Tackmann, J.~R. Walsh and S.~Zuberi, \emph{{Factorization
  and Resummation for Dijet Invariant Mass Spectra}},
  \href{http://dx.doi.org/10.1103/PhysRevD.85.074006}{\emph{Phys.Rev.} {\bf
  D85} (2012) 074006}, [\href{http://arXiv.org/abs/1106.6047}{{\tt
  arXiv:1106.6047}}].

\bibitem{Larkoski:2014tva}
A.~J. Larkoski, I.~Moult and D.~Neill, \emph{{Toward Multi-Differential Cross
  Sections: Measuring Two Angularities on a Single Jet}},
  \href{http://dx.doi.org/10.1007/JHEP09(2014)046}{\emph{JHEP} {\bf 1409}
  (2014) 046}, [\href{http://arXiv.org/abs/1401.4458}{{\tt arXiv:1401.4458}}].

\bibitem{Procura:2014cba}
M.~Procura, W.~J. Waalewijn and L.~Zeune, \emph{{Resummation of
  Double-Differential Cross Sections and Fully-Unintegrated Parton Distribution
  Functions}}, \href{http://dx.doi.org/10.1007/JHEP02(2015)117}{\emph{JHEP}
  {\bf 1502} (2015) 117}, [\href{http://arXiv.org/abs/1410.6483}{{\tt
  arXiv:1410.6483}}].

\bibitem{Larkoski:2015zka}
A.~J. Larkoski, I.~Moult and D.~Neill, \emph{{Non-Global Logarithms,
  Factorization, and the Soft Substructure of Jets}},
  \href{http://dx.doi.org/10.1007/JHEP09(2015)143}{\emph{JHEP} {\bf 09} (2015)
  143}, [\href{http://arXiv.org/abs/1501.04596}{{\tt arXiv:1501.04596}}].

\bibitem{Pietrulewicz:2016nwo}
P.~Pietrulewicz, F.~J. Tackmann and W.~J. Waalewijn, \emph{{Factorization and
  Resummation for Generic Hierarchies between Jets}},
  \href{http://dx.doi.org/10.1007/JHEP08(2016)002}{\emph{JHEP} {\bf 08} (2016)
  002}, [\href{http://arXiv.org/abs/1601.05088}{{\tt arXiv:1601.05088}}].

\bibitem{Chien:2015cka}
Y.-T. Chien, A.~Hornig and C.~Lee, \emph{{Soft-collinear mode for jet cross
  sections in soft collinear effective theory}},
  \href{http://dx.doi.org/10.1103/PhysRevD.93.014033}{\emph{Phys. Rev.} {\bf
  D93} (2016) 014033}, [\href{http://arXiv.org/abs/1509.04287}{{\tt
  arXiv:1509.04287}}].

\bibitem{Collins:1981uk}
J.~C. Collins and D.~E. Soper, \emph{{Back-To-Back Jets in QCD}},
  \href{http://dx.doi.org/10.1016/0550-3213(81)90339-4}{\emph{Nucl. Phys.} {\bf
  B193} (1981) 381}. [Erratum: Nucl. Phys.B213,545(1983)].

\bibitem{Collins:1984kg}
J.~C. Collins, D.~E. Soper and G.~F. Sterman, \emph{{Transverse Momentum
  Distribution in Drell-Yan Pair and W and Z Boson Production}},
  \href{http://dx.doi.org/10.1016/0550-3213(85)90479-1}{\emph{Nucl. Phys.} {\bf
  B250} (1985) 199--224}.

\bibitem{Collins:1985ue}
J.~C. Collins, D.~E. Soper and G.~F. Sterman, \emph{{Factorization for Short
  Distance Hadron - Hadron Scattering}},
  \href{http://dx.doi.org/10.1016/0550-3213(85)90565-6}{\emph{Nucl. Phys.} {\bf
  B261} (1985) 104--142}.

\bibitem{Collins:1988ig}
J.~C. Collins, D.~E. Soper and G.~F. Sterman, \emph{{Soft Gluons and
  Factorization}},
  \href{http://dx.doi.org/10.1016/0550-3213(88)90130-7}{\emph{Nucl. Phys.} {\bf
  B308} (1988) 833--856}.

\bibitem{Collins:1989gx}
J.~C. Collins, D.~E. Soper and G.~F. Sterman, \emph{{Factorization of Hard
  Processes in QCD}},
  \href{http://dx.doi.org/10.1142/9789814503266_0001}{\emph{Adv. Ser. Direct.
  High Energy Phys.} {\bf 5} (1989) 1--91},
  [\href{http://arXiv.org/abs/hep-ph/0409313}{{\tt hep-ph/0409313}}].

\bibitem{Moult:2016cvt}
I.~Moult, L.~Necib and J.~Thaler, \emph{{New Angles on Energy Correlation
  Functions}}, \href{http://dx.doi.org/10.1007/JHEP12(2016)153}{\emph{JHEP}
  {\bf 12} (2016) 153}, [\href{http://arXiv.org/abs/1609.07483}{{\tt
  arXiv:1609.07483}}].

\bibitem{Korchemsky:1999kt}
G.~P. Korchemsky and G.~F. Sterman, \emph{{Power corrections to event shapes
  and factorization}},
  \href{http://dx.doi.org/10.1016/S0550-3213(99)00308-9}{\emph{Nucl.Phys.} {\bf
  B555} (1999) 335--351}, [\href{http://arXiv.org/abs/hep-ph/9902341}{{\tt
  hep-ph/9902341}}].

\bibitem{Korchemsky:2000kp}
G.~Korchemsky and S.~Tafat, \emph{{On power corrections to the event shape
  distributions in QCD}},
  \href{http://dx.doi.org/10.1088/1126-6708/2000/10/010}{\emph{JHEP} {\bf 0010}
  (2000) 010}, [\href{http://arXiv.org/abs/hep-ph/0007005}{{\tt
  hep-ph/0007005}}].

\bibitem{Hoang:2007vb}
A.~H. Hoang and I.~W. Stewart, \emph{{Designing gapped soft functions for jet
  production}},
  \href{http://dx.doi.org/10.1016/j.physletb.2008.01.040}{\emph{Phys.Lett.}
  {\bf B660} (2008) 483--493}, [\href{http://arXiv.org/abs/0709.3519}{{\tt
  arXiv:0709.3519}}].

\bibitem{Ligeti:2008ac}
Z.~Ligeti, I.~W. Stewart and F.~J. Tackmann, \emph{{Treating the b quark
  distribution function with reliable uncertainties}},
  \href{http://dx.doi.org/10.1103/PhysRevD.78.114014}{\emph{Phys.Rev.} {\bf
  D78} (2008) 114014}, [\href{http://arXiv.org/abs/0807.1926}{{\tt
  arXiv:0807.1926}}].

\bibitem{Larkoski:2013eya}
A.~J. Larkoski, G.~P. Salam and J.~Thaler, \emph{{Energy Correlation Functions
  for Jet Substructure}},
  \href{http://dx.doi.org/10.1007/JHEP06(2013)108}{\emph{JHEP} {\bf 1306}
  (2013) 108}, [\href{http://arXiv.org/abs/1305.0007}{{\tt arXiv:1305.0007}}].

\bibitem{Sveshnikov:1995vi}
N.~Sveshnikov and F.~Tkachov, \emph{{Jets and quantum field theory}},
  \href{http://dx.doi.org/10.1016/0370-2693(96)00558-8}{\emph{Phys.Lett.} {\bf
  B382} (1996) 403--408}, [\href{http://arXiv.org/abs/hep-ph/9512370}{{\tt
  hep-ph/9512370}}].

\bibitem{Korchemsky:1997sy}
G.~P. Korchemsky, G.~Oderda and G.~F. Sterman, \emph{{Power corrections and
  nonlocal operators}}, \href{http://dx.doi.org/10.1063/1.53732}{\emph{AIP
  Conf.Proc.} {\bf 407} (1997) 988},
  [\href{http://arXiv.org/abs/hep-ph/9708346}{{\tt hep-ph/9708346}}].

\bibitem{Lee:2006nr}
C.~Lee and G.~F. Sterman, \emph{{Momentum Flow Correlations from Event Shapes:
  Factorized Soft Gluons and Soft-Collinear Effective Theory}},
  \href{http://dx.doi.org/10.1103/PhysRevD.75.014022}{\emph{Phys.Rev.} {\bf
  D75} (2007) 014022}, [\href{http://arXiv.org/abs/hep-ph/0611061}{{\tt
  hep-ph/0611061}}].

\bibitem{Bauer:2008dt}
C.~W. Bauer, S.~P. Fleming, C.~Lee and G.~F. Sterman, \emph{{Factorization of
  e+e- Event Shape Distributions with Hadronic Final States in Soft Collinear
  Effective Theory}},
  \href{http://dx.doi.org/10.1103/PhysRevD.78.034027}{\emph{Phys.Rev.} {\bf
  D78} (2008) 034027}, [\href{http://arXiv.org/abs/0801.4569}{{\tt
  arXiv:0801.4569}}].

\bibitem{Abbate:2010xh}
R.~Abbate, M.~Fickinger, A.~H. Hoang, V.~Mateu and I.~W. Stewart, \emph{{Thrust
  at $N^3LL$ with Power Corrections and a Precision Global Fit for
  alphas(mZ)}},
  \href{http://dx.doi.org/10.1103/PhysRevD.83.074021}{\emph{Phys.Rev.} {\bf
  D83} (2011) 074021}, [\href{http://arXiv.org/abs/1006.3080}{{\tt
  arXiv:1006.3080}}].

\bibitem{Stewart:2014nna}
I.~W. Stewart, F.~J. Tackmann and W.~J. Waalewijn, \emph{{Dissecting Soft
  Radiation with Factorization}},
  \href{http://dx.doi.org/10.1103/PhysRevLett.114.092001}{\emph{Phys.Rev.Lett.}
  {\bf 114} (2015) 092001}, [\href{http://arXiv.org/abs/1405.6722}{{\tt
  arXiv:1405.6722}}].

\bibitem{Hoang:2014wka}
A.~H. Hoang, D.~W. Kolodrubetz, V.~Mateu and I.~W. Stewart,
  \emph{{$C$-parameter distribution at N$^3$LL′ including power
  corrections}},
  \href{http://dx.doi.org/10.1103/PhysRevD.91.094017}{\emph{Phys.Rev.} {\bf
  D91} (2015) 094017}, [\href{http://arXiv.org/abs/1411.6633}{{\tt
  arXiv:1411.6633}}].

\bibitem{Lee:2007jr}
C.~Lee, \emph{{Universal nonperturbative effects in event shapes from
  soft-collinear effective theory}},
  \href{http://dx.doi.org/10.1142/S021773230702289X}{\emph{Mod.Phys.Lett.} {\bf
  A22} (2007) 835--851}, [\href{http://arXiv.org/abs/hep-ph/0703030}{{\tt
  hep-ph/0703030}}].

\bibitem{Lee:2006fn}
C.~Lee and G.~F. Sterman, \emph{{Universality of nonperturbative effects in
  event shapes}}, {\emph{eConf} {\bf C0601121} (2006) A001},
  [\href{http://arXiv.org/abs/hep-ph/0603066}{{\tt hep-ph/0603066}}].

\bibitem{Abbate:2012jh}
R.~Abbate, M.~Fickinger, A.~H. Hoang, V.~Mateu and I.~W. Stewart,
  \emph{{Precision Thrust Cumulant Moments at $N^3$LL}},
  \href{http://dx.doi.org/10.1103/PhysRevD.86.094002}{\emph{Phys.Rev.} {\bf
  D86} (2012) 094002}, [\href{http://arXiv.org/abs/1204.5746}{{\tt
  arXiv:1204.5746}}].

\bibitem{Sjostrand:2006za}
T.~Sjostrand, S.~Mrenna and P.~Z. Skands, \emph{{PYTHIA 6.4 Physics and
  Manual}}, \href{http://dx.doi.org/10.1088/1126-6708/2006/05/026}{\emph{JHEP}
  {\bf 0605} (2006) 026}, [\href{http://arXiv.org/abs/hep-ph/0603175}{{\tt
  hep-ph/0603175}}].

\bibitem{Sjostrand:2014zea}
\emph{{An Introduction to PYTHIA 8.2}},
  \href{http://dx.doi.org/10.1016/j.cpc.2015.01.024}{\emph{Comput. Phys.
  Commun.} {\bf 191} (2015) 159--177},
  [\href{http://arXiv.org/abs/1410.3012}{{\tt arXiv:1410.3012}}].

\bibitem{Cacciari:2011ma}
M.~Cacciari, G.~P. Salam and G.~Soyez, \emph{{FastJet User Manual}},
  \href{http://dx.doi.org/10.1140/epjc/s10052-012-1896-2}{\emph{Eur.Phys.J.}
  {\bf C72} (2012) 1896}, [\href{http://arXiv.org/abs/1111.6097}{{\tt
  arXiv:1111.6097}}].

\bibitem{Cacciari:2008gp}
M.~Cacciari, G.~P. Salam and G.~Soyez, \emph{{The Anti-k(t) jet clustering
  algorithm}},
  \href{http://dx.doi.org/10.1088/1126-6708/2008/04/063}{\emph{JHEP} {\bf 04}
  (2008) 063}, [\href{http://arXiv.org/abs/0802.1189}{{\tt arXiv:0802.1189}}].

\bibitem{Dokshitzer:1997in}
Y.~L. Dokshitzer, G.~Leder, S.~Moretti and B.~Webber, \emph{{Better jet
  clustering algorithms}}, {\emph{JHEP} {\bf 9708} (1997) 001},
  [\href{http://arXiv.org/abs/hep-ph/9707323}{{\tt hep-ph/9707323}}].

\bibitem{Wobisch:1998wt}
M.~Wobisch and T.~Wengler, \emph{{Hadronization corrections to jet
  cross-sections in deep inelastic scattering}},  in \emph{{Monte Carlo
  generators for HERA physics. Proceedings, Workshop, Hamburg, Germany,
  1998-1999}}, pp.~270--279, 1998.
\newblock \href{http://arXiv.org/abs/hep-ph/9907280}{{\tt hep-ph/9907280}}.

\bibitem{Wobisch:2000dk}
M.~Wobisch, \emph{{Measurement and QCD analysis of jet cross-sections in deep
  inelastic positron proton collisions at $\sqrt{s} = 300$~GeV}},  2000.

\bibitem{fjcontrib}
``Fastjet contrib.'' \url{http://fastjet.hepforge.org/contrib/}.

\end{thebibliography}\endgroup

\end{document}